\documentclass{aa}  
\usepackage{natbib}
\bibpunct{(}{)}{;}{a}{}{,} 
\usepackage{graphicx}
\usepackage{adjustbox} 
\usepackage{subcaption}
\usepackage{verbatim}
\usepackage[varg]{txfonts}
\usepackage[section]{placeins} 
\usepackage{setspace}

\usepackage{booktabs}
\usepackage{multirow}
\usepackage{dblfloatfix} 
\usepackage{comment}
\usepackage{float}
\usepackage{tikz}
\usepackage{color}
\graphicspath{{./}{./figure/}}

\usepackage{xcolor}

\usepackage[colorlinks=true, linkcolor=blue, citecolor=blue, filecolor=blue, urlcolor=blue]{hyperref}

\usetikzlibrary{arrows,positioning} 
\tikzset{
    >=stealth',
    punkt/.style={
           rectangle,
           rounded corners,
           draw=black, very thick,
           text width=6.5em,
           minimum height=2em,
           text centered},
    pil/.style={
           ->,
           thick,
           shorten <=2pt,
           shorten >=2pt,}
}

\newcommand\rectagular[1][red]{\begin{tikzpicture}
\draw [fill=red,red] (0.2,0.2) rectangle (0.3,0.3); 
\end{tikzpicture}
}

\begin{document}

    \title{Ensemble modeling of Coronal Mass Ejection dynamics and forecasts at 1~AU with a semi-analytic flux-rope model}
    \author{S.Stamkos\inst{1}
    \and S.Patsourakos\inst{1} 
    \and A. Vourlidas\inst{2}
    \and E. Paouris\inst{2}
  }
\institute{Department of Physics, Section of Astrogeophysics, University of Ioannina, 45110, Greece. 
\and Applied Physics Laboratory, Johns Hopkins University, Laurel, MD 20723, USA
}
    
\date{Received date / Accepted date }

\abstract {} {This study quantifies how uncertainty in physically meaningful coronal mass ejection (CME) and solar-wind inputs propagates into forecast-relevant diagnostics from eruption through heliospheric evolution. Using a semi-analytic erupting flux rope (EFR) model, we evaluate how initial eruption parameters (spatial scales, density structure, and magnetic properties) together with background solar-wind conditions shape the distributions of the arrival time, kinematics, magnetic-field measures, and impact duration at 1~AU.}{We used a semi-analytic flux-rope model to simulate CME initiation and Sun-to-1~AU propagation. The model includes Lorentz, gravitational, and drag forces and is driven by a prescribed, time-dependent poloidal-flux injection. Relative to the original EFR formulation, we incorporated sheath and pile-up effects via an effective (virtual) mass and updated the drag term for the CME–solar-wind coupling. We embedded the model in a Monte Carlo scheme with truncated-normal sampling of key inputs to quantify dispersion in the arrival time, kinematics, internal magnetic field, and impact duration at 1~AU.}{Across six CME events, the ensembles show moderate event-dependent dispersion in the 1~AU diagnostics. For the $\pm 20\%$ sampling, all reported spreads correspond to the 1$\sigma$ standard deviation across the ensemble. The ensemble time-of-arrival (ToA) spread is 2.4--7.7~h across the six events, and it is primarily controlled by the poloidal-flux injection history (injected-flux amplitude as well as rise and plateau timing) together with CME--solar-wind coupling (upstream wind speed and drag coefficient), with the top-ranked contributor shifting between events. The leading-edge speed shows a spread of 28--53~km~s$^{-1}$ and is primarily controlled by background-flow properties, while injection-related terms act as secondary contributors in all cases. The magnetic-field diagnostics demonstrate two behaviors: the sheath field is relatively tightly distributed, with a spread of 1--3.5~nT, and is mainly controlled by upstream solar-wind conditions together with global size and expansion scaling, whereas the internal flux-rope field shows a larger spread of 1--7.6~nT and is primarily governed by eruption-driving and flux-content parameters. The spread in the duration of impact spans 2.4--6.3~h and is mainly controlled by geometric size and expansion scaling, with additional sensitivity to the temporal characteristics of the driving.}{Embedding the EFR model in a Monte Carlo framework allows us to quantify how uncertainty in its inputs translates into dispersion in Sun-to-1~AU kinematics and in key 1~AU diagnostics for the events studied here. Within this framework, the feature-ranking analysis identifies which EFR inputs most strongly drive that dispersion, yielding a clear ordering of the most important parameters as event-specific information improves. In operational applications or focused event studies, the same workflow can be rerun with progressively narrower input ranges as additional constraints are obtained, yielding correspondingly tighter and more defensible confidence bounds on 1~AU arrival and impact diagnostics.}


\titlerunning{Ensemble Modeling of CME Dynamics}

\authorrunning {S.Stamkos et al.}

\maketitle
\nolinenumbers

\section{Introduction} \label{introduction}

Coronal mass ejections (CMEs) are among the most powerful phenomena originating from the Sun. They expel over $10^{15}$ g of magnetized plasma into the heliosphere at speeds ranging from a few hundred to more than 2000 km s$^{-1}$, generating kinetic energies on the order of $10^{32}$--$10^{33}$ erg \citep{2012LRSP....9....3W}. When directed toward Earth, CMEs can trigger severe geomagnetic storms that disrupt satellite operations, pose radiation risks to astronauts, and jeopardize ground-based infrastructures, including power grids and communication systems \citep{2013SpWea..11..585B, 2015SpWea..13..734P}. As modern society increasingly depends on space infrastructure and GNSS-dependent services, accurate and timely space-weather forecasting has become both an operational necessity and a scientific priority \citep{2015AdSpR..55.2745S, 2017RiskA..37..206E, 2014NatCo...5.3481L}.

At the heart of this challenge lies the need to understand the behavior of CMEs from their initiation at the Sun to their evolution through the heliosphere and the interaction with the Earth’s magnetosphere. Progress is limited by the complex, coupled magnetic, and plasma processes that govern CME propagation and by sparse constraints on key coronal and heliospheric conditions, motivating the continued development and evaluation of predictive models.

Central to this goal is the ability to predict how CMEs evolve as they propagate through the highly variable and often turbulent heliospheric environment. Missions such as SOHO, Hinode, STEREO, SDO, Parker Solar Probe, and Solar Orbiter have significantly enhanced our observational toolkit, providing valuable insights into CME initiation and large-scale propagation characteristics. However, despite these advances, accurately forecasting CME arrival times remains an enduring challenge, largely because CMEs interact in complex ways with the ambient solar wind during transit \citep{2019RSPTA.37780096V}. Assessing geoeffectiveness is even more difficult because it depends on how a CME's internal magnetic structure evolves during propagation and on the magnetic configuration that ultimately reaches 1~AU. In practice, the most relevant quantities include the arrival time and leading-edge speed, the magnetic-field strength carried by the ejecta and the sheath (and its orientation, especially any sustained southward component), and the impact duration, which together govern the strength and persistence of the CME—magnetosphere coupling \citep{2019RSPTA.37780096V, 2007JGRA..11210102Z}.

A wide array of modeling strategies have been developed to predict CME propagation and arrival. Full three-dimensional magnetohydrodynamic (MHD) simulations provide the most physically self-consistent description of CME--solar-wind interaction, but they remain computationally demanding, and their performance can be strongly affected by uncertain or poorly constrained inputs, including boundary conditions or coronal prescriptions \citep{2003ApJ...595L..57R, 2004JGRA..109.2107M, 2005JGRA..11012226T, 2017SSRv..212.1159M, 2004JGRA..109.2116O}. 

Empirical or reduced-physics models (e.g., drag-based and cone-type formulations) offer computational efficiency and useful arrival-time skill. However, they typically rely on simplified geometries and reduced dimensionality and therefore capture only a subset of the coupled processes that shape CME evolution, with limited ability to represent internal structure \citep{2013SoPh..285..295V, 2015SoPh..290.1775M}.

Ensemble forecasting has become central to translating uncertain CME and solar-wind inputs into probabilistic outcomes. Operational ensemble prediction with the Wang-Sheeley-Arge (WSA) model coupled to ENLIL demonstrates this explicitly: even when an ensemble is summarized by a single representative forecast (e.g., the ensemble mean), typical time-of-arrival (ToA) errors remain of the order of $\sim$10~h, and biases (particularly for fast CMEs) can be traced to combinations of uncertain input speed and angular width \citep{2015SoPh..290.1775M}. Community benchmarking through the NASA Community Coordinated Modeling Center (CCMC) CME Scoreboard shows that, across widely used arrival-time models, mean performance remains at the $\sim$10~h level with substantial event-to-event scatter, and that a simple multi-model ``average of all methods'' forecast performs competitively \citep{2018SpWea..16.1245R}. Consistent with this, mean absolute arrival-time errors reported across widely used models are typically $\sim$10--13~h \citep{2019RSPTA.37780096V}, and an updated analysis of CME Scoreboard arrival-time predictions reports an overall mean absolute error (MAE) of 13.2~h \citep{2024SpWea..2203951K}.  Heliospheric-imaging ensemble studies reach a consistent conclusion: varying plausible model setups and input choices produces large differences in prediction accuracy, and accurate specification of the ambient solar wind emerges as a key control on arrival predictions \citep{2021SpWea..1902553A}. Together, these results motivate uncertainty-aware CME modeling in which both CME and background-wind uncertainties are treated explicitly, and the ensemble spread is interpreted as a probabilistic forecast rather than a single best-guess trajectory.

Semi-analytic models provide an intermediate approach, aiming to capture the dominant forces governing CME dynamics while retaining physical transparency and computational efficiency. These models idealize the CME as a simplified magnetic structure, usually a flux rope, and solve for its motion under the influence of magnetic, gravitational, and drag forces. Notable examples include the toroidal instability model \citep{2006PhRvL..96y5002K}, catastrophe models \citep{1991ApJ...373..294F}, and the erupting flux rope (EFR) model \citep{1989ApJ...338..453C, 1996JGR...10127499C, 1993GeoRL..20.2319C}. Such frameworks are particularly valuable for comparative studies across events because they clarify cause–effect relationships by linking internal parameters (e.g., current, mass loading, and footpoint separation) to observable kinematics.

Among these, the EFR model of \citet{1989ApJ...338..453C} and subsequent developments \citep{1993GeoRL..20.2319C, 1996JGR...10127499C, 2010ApJ...717.1105C} have been in a number of case studies to reproduce impulsive and gradual acceleration profiles through a physically motivated prescription for poloidal-flux injection. However, most EFR applications remain single case studies and largely deterministic. As a result, realistic input uncertainty has rarely been propagated through EFR to quantify its effect on space-weather-relevant outputs, and sheath contributions are often left out. That matters because compression and pile-up ahead of the ejecta can shape the density and field signatures near the front and can influence the inferred impact window at 1~AU.

These issues motivate the present study. Our goal is not to validate EFR as an operational forecasting tool but to use the EFR framework as a physically interpretable test to quantify how uncertainty in its inputs propagates through the model dynamics and emerges as dispersion in 1~AU diagnostics, providing a basis for defensible confidence bounds and for identifying which inputs most strongly limit predictability. To do so, we embed the EFR model in a Monte Carlo framework that perturbs key input-parameter groups and converts those perturbations into distributions of arrival time, kinematics, magnetic-field measures, and impact duration at 1~AU. We also extend the ``classic'' EFR implementation by incorporating a sheath/pile-up treatment and updating the drag term to better represent CME--solar-wind momentum coupling. Together, these elements enable a controlled assessment of (i) how compounded input uncertainty produces dispersion in forecast-relevant outputs and (ii) which EFR parameters most strongly organize that spread within the model.

The remainder of this paper is organized as follows. Section 2 describes the EFR model used in this work, including its conceptual basis, flux-rope geometry and plasma properties, sheath formation, the forces governing propagation, and the poloidal-flux injection prescription. Section 3 describes the Monte Carlo simulation framework and the sampling strategy used to quantify uncertainty and parameter sensitivity. Section 4 presents the results of the ensemble experiments, six CME events (23 February 1997, 30 April 1997, 2 June 1998, 11 June 1998, 17 May 2008, 3 April 2010), with emphasis on ToA statistics and associated geoeffective-relevant quantities at 1~AU. Section 5 discusses the implications of these findings, the limitations of the present approach, and the priorities for future improvements. Finally, Section 6 summarizes the main conclusions and discusses extensions
of the EFR model, allowing for better constraining of its input parameters, by harnessing STEREO and SDO observations.

\section{The erupting flux rope model}

\subsection{Conceptual framework}

Initially developed by \citet{1989ApJ...338..453C}, the EFR model treats the CME as a toroidal magnetic flux rope with anchored footpoints and evaluates its motion under the influence of the Lorentz self-force, gravitational pull, and aerodynamic drag. A central element of the EFR model is the role of the hoop force, a manifestation of the Lorentz force due to the curvature of the poloidal field, which drives the upward acceleration when the rope’s equilibrium is disrupted. This force, originally developed in the context of tokamak plasma physics \citep{1966RvPP....2..103S}, was adapted to the solar context by accounting for the footpoint anchoring and non-axisymmetric flux rope geometry \citep{1993GeoRL..20.2319C}. The EFR model reconstructs CME kinematics with high fidelity by employing a minimal set of physically grounded parameters. Unlike full time-dependent MHD simulations, which require extensive computational resources and detailed boundary conditions, the EFR framework provides an analytically manageable alternative that preserves essential physical realism.

A key feature of the model is its ability to reproduce both impulsive and gradual CME acceleration profiles through variation in a single parameter: the time-dependent injection of poloidal magnetic flux, $d\Phi_p/dt$. In the EFR formulation, this process increases the poloidal magnetic field within the rope, amplifying the hoop force and thereby driving the eruption \citep{2006ApJ...649..452C, 2010ApJ...717.1105C}. In this sense, the triggering and driving are embedded in the flux-rope dynamics through the prescribed $\Phi_p(t)$ history.

Event-based EFR studies have inferred $d\Phi_p(t)/dt$ by fitting modeled height--time profiles to coronagraph observations and then comparing the resulting injection profiles with the associated soft X-ray emission. For the events analyzed by \citet{2010ApJ...717.1105C}, the best fit $d\Phi_p(t)/dt$ profiles exhibit close temporal correspondence with the SXR light curves (e.g., comparable durations and similar rise/decay behavior), which the authors interpret as evidence for a linkage between the injection history, flare-energy release, and CME acceleration.

The model naturally captures the multiphase kinematic evolution reported in coronagraph observations of CME onset. In particular, \citet{2001ApJ...559..452Z} identified an impulsive acceleration phase in the low corona that coincides with the flare rise phase, followed by a propagation phase characterized by an approximately constant speed or a slowly varying speed. In the flux-rope framework, \citet{2006ApJ...649..452C} similarly distinguish a Lorentz-dominated main-acceleration regime in the inner corona from a residual-acceleration regime at larger heights, where Lorentz driving weakens and competing forces such as gravity and drag become increasingly important. The transition height is event-dependent and scales with the flux-rope size. Synthetic coronagraph images generated from flux-rope models reproduce key CME morphologies, including both halo and limb events \citep{2000ApJ...539..964K}.

Furthermore, the relative simplicity and physical transparency of the model make it useful for case study diagnostics. By fitting observed CME height–time measurements with EFR solutions, one can constrain the time profile of poloidal-flux injection, $d\Phi_p/dt$, and derive quantities such as the total injected poloidal flux and the associated increase in poloidal magnetic energy \citep{2010ApJ...717.1105C}. This diagnostic approach has also been extended to heliospheric distances: for a CME continuously tracked by SECCHI (Sun–Earth Connection Coronal and Heliospheric Investigation) from eruption to 1~AU and subsequently sampled in situ, an EFR solution constrained by the remote-sensing trajectory alone reproduced the kinematics to within $\sim$1\% over the 1~AU field of view and yielded magnetic-field and plasma properties at 1~AU consistent with the in situ measurements \citep{2010ApJ...715L..80K}. Together with the flux-rope scaling that links the height of peak acceleration to the footpoint separation scale, these results motivate the broader use of flux-rope models incorporating Lorentz forces and flux injection as components of more physics-based forecasting frameworks when constrained by routine coronagraph and EUV observations (e.g., from SOHO and STEREO). However, their integration into operational systems remains limited, in part because key inputs are not yet straightforward to extract automatically and robustly from remote-sensing data \citep{2006ApJ...649..452C}.

Together, these results indicate that although the EFR model captures CME initiation and acceleration and has been successfully tested against observations, a more complete picture of CME evolution also requires careful treatment of the internal magnetic and plasma structure of the flux rope. We therefore turn to a concise description of these properties.

\subsection{Geometrical and plasma properties of the flux rope}

The flux rope comprises two principal magnetic components: the toroidal and poloidal magnetic fields, denoted $B_t$ and $B_p$ respectively. The toroidal field, generated by a poloidal current $J_p$, runs along the axis of the rope, while the poloidal field, arising from a toroidal current $J_t$, encircles the axis in the local azimuthal direction \citep{1989ApJ...338..453C, 2001ApJ...562.1045K}. This magnetic configuration yields a twisted field topology that confines plasma and defines the rope’s mechanical properties. The major radius $R$ describes the curvature of the rope as it arches through the corona, while the minor radius $a$ characterizes the thickness of the current channel. The footpoint separation $S_f$, anchored at the photosphere, fixes the global geometry and plays a key role in the subsequent dynamics \citep{2003JGRA..108.1410C, 2006ApJ...649..452C}.

Anchoring in the dense solar photosphere serves as an effective boundary condition preventing translational motion of the footpoints during the initial phase of the eruption. This anchoring is often modeled as rigid line-tied boundary conditions in both analytic and numerical treatments \citep{2003JGRA..108.1410C, 2000ApJ...539..964K}. The curvature of the flux rope then gives rise to the outward-directed Lorentz self-force (the hoop force introduced in 2.4), which dominates the initial acceleration.

Further refinements to the flux-rope structure include a cold, dense component (prominence plasma) embedded within a hot, lower-density cavity \citep{1996JGR...10127499C}. This multicomponent picture is consistent with coronagraph observations in which prominence material trails the bright CME front within the rope’s trailing edge \citep{1986JGR....9110951I}. Although the cold component may drain or dissipate as the eruption proceeds, its initial presence modifies the mass distribution, drag profile, and internal pressure gradients, and it helps explain the classic three-part morphology and reconcile modeled vs.\ measured densities and magnetic fields at 1~AU \citep{1996JGR...10127499C, 1993GeoRL..20.2319C, 2001ApJ...562.1045K}.

To capture the evolving geometry, we model the flux rope as a toroidal loop with time-dependent height $z$, minor radius $a$, and major radius $R$, line-tied at the solar surface with fixed footpoint separation $S_f$. The major radius follows the geometric constraint
\begin{equation}
R = \frac{z^2 + \left(S_f / 2\right)^2}{2z},
\end{equation}
the angular extent is
\begin{equation}
\theta = \arcsin\!\left(\frac{S_f}{2R}\right),
\end{equation}
and the arc length $L$ (the current-channel length) is
\begin{equation}
L = \begin{cases}
2\pi R\,\dfrac{\theta}{\pi}, & z < S_f / 2, \\
2\pi R\!\left(1 - \dfrac{\theta}{\pi}\right), & \text{otherwise},
\end{cases}
\end{equation}
which accommodates both upward- and downward-concave configurations depending on the apex position.

The total CME mass is obtained by integrating over the rope volume and accounting for both the cavity plasma and embedded prominence material,
\begin{equation}
M_{\rm CME} = \pi a^2 L \cdot m_p \cdot 1.1 \cdot \big(n_{\rm cavity} + n_{\rm prominence}\big),
\end{equation}
where $m_p$ is the proton mass and the factor $1.1$ accounts for a $10\%$ helium abundance. The prominence density is treated as a prescribed, time-dependent enhancement of the cavity density. At early times (during a short post-onset relaxation interval), the prominence is assigned a constant density contrast relative to the cavity, ensuring an initially mass-loaded core. After this relaxation period, the enhancement is allowed to diminish smoothly on a characteristic timescale, representing the progressive loss of prominence material through heating/dissipation and expansion, while approaching a finite residual level so that the prominence contribution does not vanish entirely.

\subsection{Sheath formation, pile-up, and Rankine--Hugoniot relations}

The CME--solar-wind interaction adds inertia to the propagating structure because an accelerating body must displace and accelerate surrounding fluid. In fluid dynamics this is commonly described as added (or virtual) mass \citep{white2011fluid}. In the CME context, the same physical idea applies: part of the ambient plasma is entrained and accelerated in the CME's vicinity, and mass accumulation ahead of the ejecta contributes to the compressed pile-up region that ultimately forms the sheath \citep{2004SoPh..221..135C,2021JSWSC..11...34V}.

Accordingly, we write the effective mass of the CME--solar-wind system as
\begin{equation}\label{eq:mtot}
M_{\mathrm{tot}} = M_{\mathrm{CME}} + M_{\mathrm{virtual}},
\end{equation}
where $M_{\mathrm{CME}}$ is the intrinsic flux-rope mass. The added-mass term represents the inertia of displaced ambient solar-wind plasma. In our implementation, we approximate this contribution as a constant-coefficient fraction of the solar-wind mass contained in the instantaneous CME volume,
\begin{equation}\label{eq:mvirtual}
M_{\rm virtual} = \frac{1}{2}\,\pi a^2 L \cdot m_p \cdot n_{\rm sw},
\end{equation}
with the coefficient $1/2$ adopting a standard added-mass scaling for a body moving in a surrounding fluid \citep{book:1197664} and $n_{\rm sw}$ evaluated from the background model. This term enters the EFR equation of motion through the effective inertia $M_{\mathrm{tot}}$, and becomes small when the CME density greatly exceeds the ambient density (i.e., when $M_{\mathrm{CME}} \gg M_{\mathrm{virtual}}$), a limit in which added-mass effects are frequently neglected in drag-type applications \citep{2013SoPh..285..295V}.

Beyond the near-Sun regime, the intrinsic CME mass is commonly assumed to approach an approximately constant value. In that case, the distance dependence of $M_{\mathrm{tot}}$ is governed mainly by the evolving ambient density and the expanding CME volume, so the effective inertia can vary as the CME propagates. This motivates treating CME propagation as a variable-inertia problem when interpreting the role of pile-up and sheath formation in the momentum budget.

When the CME exceeds the local fast--magnetosonic speed, a shock forms ahead of its leading edge and the region between the shock and the CME front becomes the sheath. This sheath contains shock-processed solar wind that is compressed and heated, with enhanced magnetic fields and elevated thermal and dynamic pressures \citep{2017LRSP...14....5K}. The macroscopic state across the shock is set by the MHD Rankine--Hugoniot relations, which connect upstream and downstream quantities through control parameters including the Alfv\'en Mach number, shock obliquity (the angle between the upstream field and the shock normal), and the adopted polytropic index \citep{2022SpWea..2003165K}. Operationally, we compute the density compression from the MHD jump conditions and then determine the downstream magnetic field and pressure self-consistently, retaining the essential dependence on shock geometry and flow strength without reproducing the full algebra. Physically, larger Alfv\'en Mach numbers and more perpendicular shocks yield stronger compression and heating, whereas quasi-parallel configurations produce more modest amplification.

The sheath's radial extent reflects a competition between kinematics and compression: it broadens as CME bulk motion (and expansion) increasingly outpaces the ambient solar wind, while for fixed kinematics stronger compression tends to confine the layer. Framed this way, the sheath is a dynamically regulated interface whose magnetic amplification and pressure enhancement are central to geoeffectiveness.

\subsection{The forces driving CME propagation}

Building on the conceptual description in Section~2.1, the CME dynamics are fundamentally governed by the interplay of macroscopic forces acting on the erupting magnetic flux rope. Chief among these is the aforementioned Lorentz hoop force, which arises from the internal toroidal current interacting with its own poloidal magnetic field. This force is directed radially outward and acts to expand and lift the toroidal flux rope when equilibrium is perturbed. Conceptually, the flux rope behaves as a magnetic spring: it stores energy in its twisted field lines and releases this energy dynamically during eruption \citep{1966RvPP....2..103S, 1989ApJ...338..453C}. Because the hoop force is intrinsic to the magnetic structure and does not require coupling to the ambient plasma, it is the primary agent for rapid CME acceleration in the inner corona \citep{1996JGR...10127499C, 2000ApJ...539..964K}.

To properly quantify the dynamics in realistic flux-rope configurations, we use the full expression for the Lorentz acceleration, which includes curvature, internal pressure/tension, external magnetic pressure, and inductive feedback,
\begin{equation}
\label{eq:lorentz}
    F_{L} = \frac{I_t^2}{c^2 M_{total} R} \left[ \ln\left(\frac{8R}{a}\right) + \frac{1}{2} \beta_p - \frac{1}{2} \left(\frac{\bar{B}_t^2}{\bar{B}_p^2} \right) + 2 \left( \frac{R}{a} \right) \left( \frac{B_{\mathrm{ext}}}{\bar{B}_p} \right) - 1 + \frac{\xi}{2} \right],
\end{equation}
where $I_t$ is the toroidal current, $\bar{B}_p$ and $\bar{B}_t$ are the average poloidal and toroidal magnetic field components respectively, $B_{\mathrm{ext}}$ is the ambient coronal magnetic field, and $R$ and $a$ are the major and minor radii. The parameters $\beta_p$ and $\xi$ capture, respectively, the contribution of plasma pressure relative to the poloidal field and the dependence on the internal poloidal-field profile. We do not reproduce their standard definitions here, as the above form suffices for our modeling \citep{1993GeoRL..20.2319C, 2010ApJ...717.1105C, 2012PhDT.......120K}. For context, the commonly used thin-rope approximation for the hoop force per unit length, which scales as $I_t^2\,\ln(8R/a)/(c^2 R)$, is omitted here since the full acceleration already encapsulates this dependence.

The Lorentz term above is obtained from an energy-based formulation in which the evolving self-inductance of the toroidal rope feeds back on the current. Explicit forms for the self-inductance and current evolution (e.g., $L_{\rm in}(a,R)$ and $I_t$ as functions of geometry and poloidal flux) are standard and not repeated here. We simply note that the $R$- and $a$-dependences entering $F_L$ arise from that inductive coupling \citep{2000ApJ...539..964K}.

The total effective mass must include both the CME’s own content and the virtual mass due to displaced ambient plasma:
\begin{equation}
    M_{\mathrm{total}} = M_{\mathrm{prominence}} + M_{\mathrm{cavity}} + M_{\mathrm{virtual}}
\end{equation}.
This choice ensures that the modeled acceleration reflects not only the internal plasma load but also solar-wind inertia.

Opposing the Lorentz force are aerodynamic drag and gravity. The aerodynamic drag becomes increasingly important in the outer corona and heliosphere, where the flux rope interacts with the solar wind. Following \citet{2004SoPh..221..135C}, we write
\begin{equation}
\label{eq:drag}
    a_{d} = -\gamma_d\,C_d\,(v_{\mathrm{CME}} - v_{\mathrm{sw}})\,|v_{\mathrm{CME}} - v_{\mathrm{sw}}|,
\end{equation}
with
\begin{equation}
    \gamma_d = \frac{4}{\pi\,a\left[\,2\left(\frac{\rho_{\mathrm{cavity}} + \rho_{\mathrm{prominence}}}{\rho_{\mathrm{sw}}}\right) + 1\right]},
\end{equation}
where $a_{d}$ is the acceleration due to drag, $C_d$ is a dimensionless drag coefficient, $\rho_{\mathrm{cavity}}$ and $\rho_{\mathrm{prominence}}$ are the internal mass densities, and $\rho_{\mathrm{sw}}$ is the local solar-wind density. This makes explicit that, for fixed $(v_{\mathrm{CME}} - v_{\mathrm{sw}})|v_{\mathrm{CME}} - v_{\mathrm{sw}}|$, larger or denser ropes undergo smaller drag accelerations.

Gravity acts downward and is most significant at low heights, particularly for mass-loaded ropes. Rather than listing the standard $1/r^2$ form for $g(z)$, we use the resulting contribution to the equation of motion:
\begin{equation}
\label{eq:gravity}
    F_g = \frac{\pi a^2 L \left[ \rho_{\mathrm{sw}} - (\rho_{\mathrm{cavity}} + \rho_{\mathrm{prominence}}) \right] m_p \, g(z)}{M_{\mathrm{total}}},
\end{equation}
where $L$ is the rope length, $m_p$ the proton mass, and $M_{\mathrm{total}}$ the total (including virtual) mass. This form emphasizes the competition between internal loading and external buoyancy in the stratified corona.

Quantitatively, the net vertical force acting on the apex of the flux rope is the sum of the relevant contributions:
\begin{equation}
\label{eq:balance}
    F_{net} =  (F_{h} +  F_{p} +  F_{tension})  +  F_{g}  +  F_{d},
\end{equation}
where $F_h$ is the hoop (Lorentz) term, $F_p$ the pressure force, $F_g$ gravity, $F_d$ drag, and $F_{\text{tension}}$ the restraining tension from overlying fields. At early times, $F_h$ dominates and initiates the acceleration. As the flux rope rises and the background pressure and tension decrease, $F_d$ and inertial terms gradually become significant. The relative contributions are time- and height-dependent, producing the characteristic two-phase acceleration profile seen in models and observations \citep{2001ApJ...559..452Z}.

In addition to the apex motion, the EFR model treats the cross-sectional expansion self-consistently. The evolution of the minor radius $a$ is governed by the competition between toroidal magnetic pressure, poloidal magnetic tension, and plasma pressure:
\begin{equation}
\label{eq:expansion}
    M_{total}\frac{d^2 a}{dt^2} = \frac{I_t^2}{c^2 a} \left( \frac{\overline{B_t^2}}{\overline{B_p^2}} - 1 + \beta_p \right),
\end{equation}
where $\overline{B_t^2}$ and $\overline{B_p^2}$ are cross-sectional means and $\beta_p = 8\pi P / \overline{B_p^2}$. Unlike kinematic prescriptions, this first-principles form naturally reproduces observed width–height relationships \citep{1996JGR...10127499C, 2001ApJ...562.1045K, 2003JGRA..108.1410C}.

This framework, developed in the EFR model \citep{1989ApJ...338..453C}, links the acceleration profile directly to geometry, including the footpoint separation $S_f$. In particular, the peak upward acceleration occurs near $Z_{\max}$, between $Z_* = S_f/2$ and $Z_m \approx 1.5\,S_f$ \citep{2006ApJ...649..452C}. The main acceleration phase ($Z \lesssim Z_m$) is dominated by the hoop force, whereas for $Z > Z_m$ the magnetic driving weakens and drag becomes increasingly important, especially for fast CMEs in denser ambient wind \citep{2013SoPh..285..295V,2011ApJ...743..101T,2010ApJ...717.1105C}. These elements provide the physical basis for the EFR model's low-corona acceleration and its transition to interplanetary propagation.

\subsection{Poloidal flux injection}

As introduced in Section~2.1, a defining feature of the EFR model is that an otherwise stable equilibrium can be destabilized by a time-dependent increase in the poloidal magnetic flux passing through the current channel. An increase in $\Phi_p$ enhances the internal poloidal field $B_p$ and therefore the outward Lorentz self-force (hoop force) on the rope. Operationally, we treat this as a prescribed temporal profile for $\Phi_p$ (and hence $d\Phi_p/dt$) that both initiates and regulates the early acceleration phase.

A nonzero $d\Phi_p/dt$ implies a finite electromotive force (EMF) in the toroidal direction and associated inductive electric fields. Within the EFR framework, these fields have been discussed in connection with particle acceleration and flare-related energetics, and event-based EFR applications constrain the parameters of $d\Phi_p/dt$ by fitting the modeled CME height--time evolution to observations \citep{2010ApJ...717.1105C}. For brevity, we do not repeat the standard derivation relating $d\Phi_p/dt$ to the induced EMF; here it is sufficient to note that the adopted injection introduces inductive fields of the same type invoked in EFR-based interpretations of flare energization. In this study, however, we do not perform event-specific fits and instead adopt nominal injection parameters from those event-based solutions as prescribed inputs.

The physical origin of the time-dependent poloidal flux used in EFR-style models remains under active study. Proposed sources include reconnection-mediated conversion of sheared arcades into flux-rope flux \citep{1999ApJ...510..485A}, emergence of twisted subphotospheric ropes \citep{2003ApJ...589L.105F}, and photospheric twisting or shearing of line-tied footpoints. In addition, poloidal flux can be added to the erupting rope during the associated flare via coronal reconnection beneath the CME, increasing the poloidal content as the eruption proceeds. However, interpreting this increase as photospheric flux injection contemporaneous with the main acceleration phase is strongly constrained by observations.

In particular, \citet{2010ApJ...714...68S} tested the photospheric flux-injection hypothesis by comparing the energy and helicity fluxes required by flux-rope fits to CME trajectories with SOHO/MDI magnetic-field and Doppler measurements at the inferred footpoints, and found that the required photospheric signatures would be unrealistically large and are not observed. In our semi-analytic implementation, we therefore treat the prescribed poloidal-flux history as a phenomenological, boundary-driven driver that increases the hoop force and lowers the loss-of-equilibrium threshold \citep{1989ApJ...338..453C, 1993GeoRL..20.2319C}, without implying that the necessary flux and energy are injected through the photosphere during the eruption.

A related empirical result from event-based fits is that the inferred $d\Phi_p/dt$ profiles closely track the soft X-ray (SXR) light curves of the associated flares: the rise and decay of $d\Phi_p/dt$ mirror the SXR evolution, suggesting a close connection between the adopted injection history, flare energy release, and CME acceleration \citep{2010ApJ...717.1105C}. We use this correspondence qualitatively to motivate the adopted parameterization.

Following Chen \& Kunkel (2010, their Eq. 9), the injection profile is

\begin{equation}
\label{injection}
\frac{d\Phi_p}{dt}=
\begin{cases}
Q_0 \equiv \left.\dfrac{d\Phi_p}{dt}\right|_{t=0} & 0 \le t \le t_1,\\[4pt]
Q_0 + Q_1\!\left[\mathrm{sech}^2\!\left(\dfrac{t-t_2}{\tau_1}\right) - \mathrm{sech}^2\!\left(\dfrac{t_1-t_2}{\tau_1}\right)\right] & t_1 < t < t_2,\\[4pt]
Q_0 + Q_1\!\left[1 - \mathrm{sech}^2\!\left(\dfrac{t_1-t_2}{\tau_1}\right)\right] \equiv \left.\dfrac{d\Phi_p}{dt}\right|_{\max} & t_2 \le t \le t_3,\\[4pt]
\left.\dfrac{d\Phi_p}{dt}\right|_{t=t_3}\,\mathrm{sech}^2\!\left(\dfrac{t-t_2}{\tau_2}\right) & t_3 < t
\end{cases}
\end{equation}
where \(Q_0\) and \(Q_1\) are nonnegative rates (Mx\,s\(^{-1}\)), \(\tau_1\) and \(\tau_2\) are the rise/decay timescales, and \(t_1,t_2,t_3\) specify the start, peak, and end of the main injection interval. In this parameterization, \(t_2-t_1\) is the duration of the rise phase, during which the injection ramps from the background level \(Q_0\) toward its maximum value, while \(t_3-t_2\) is the duration of the plateau (pulse) during which the injection remains approximately constant at \(\left.d\Phi_p/dt\right|_{\max}\). For \(t>t_3\), the injection declines from its value at \(t_3\) and decays on the timescale \(\tau_2\). 

Poloidal flux injection also modifies the rope’s current system. In a quasi-steady inductive picture, increasing $\Phi_p$ raises the toroidal current $I_t$ through the rope’s time-dependent inductance ($\Phi_p \sim L\,I_t$), further enhancing the self-force. Once the injection subsides, the system transitions to a propagation phase with diminishing magnetic drive and increasing drag influence \citep{2006ApJ...649..452C}.

Event-to-event variability then follows from differences in the amplitude and duration of $d\Phi_p/dt$: weak or extended injections map to slow CMEs, while intense, impulsive injections generate fast events. Systematic variation of the profile parameters reproduces the observed spread of speed–height relations \citep{2003JGRA..108.1410C}. Beyond reproducing kinematics, the injection paradigm unifies CME dynamics and flare energetics: the same time profile $d\Phi_p/dt$ governs the EMF available for particle acceleration, the flare heating rate, and the mechanical work on the rope \citep{2010ApJ...717.1105C}. 

In summary, poloidal flux injection serves as the primary control parameter in the EFR model: it destabilizes the preeruption configuration and sets the timing and strength of the early CME acceleration. When constrained by kinematic fits and considered alongside flare timing, it offers a testable, observation-linked driver for CME initiation and flare coupling within the EFR framework.

\section{Monte Carlo simulation framework}

\begin{table*}[t]
\centering
\caption{Model inputs varied in the Monte Carlo ensembles, grouped by the physical role in the EFR framework.}
\label{tab:mc_inputs}
\renewcommand{\arraystretch}{1}
\setlength{\tabcolsep}{5pt}

\begin{tabular*}{\textwidth}{@{\extracolsep{\fill}} p{0.24\textwidth} p{0.16\textwidth} p{0.34\textwidth} @{}}
\hline
Group & Parameters varied & notes / description \\
\hline
Geometry / spatial scales
& $h_{0}$, $s_f$, ratio
& Eruption onset height ($h_{0}$), footpoint separation (sf), and aspect ratio (ratio) controlling flux-rope geometry and expansion scaling. \\[5pt]

Poloidal-flux injection profile
& $\tau_1,\tau_2,t_1,t_2,t_3, q_1$
& Rise/decay timescales ($\tau_1,\tau_2$), timing markers ($t_1,t_2,t_3$), and injection amplitude ($q_1$) defining the prescribed $d\Phi_p/dt$ driver. \\[7pt]

Magnetic-field initialization
& $B_p, B_t$
& Initial poloidal and toroidal field strengths setting the starting magnetic content of the rope. \\[7pt]

Plasma density / mass loading
& $n_{\mathrm{cavity}_0}$, $\delta$
& Initial cavity number density ($n_{\mathrm{cavity},0}$) and prominence fall-off / residual factor ($\delta$) controlling internal mass loading and its evolution. \\[7pt]

Solar-wind background and coupling
& $\mathrm{v}_{\mathrm{sw}_0}$, $n_{\mathrm{sw}_0}$, $C_{d}$
& Upstream solar-wind speed and density, and the drag coefficient controlling momentum exchange during propagation. \\
\hline
\end{tabular*}
\end{table*}

\begin{table*}[t]
\centering
\caption{Context summary of selected CMEs.}
\label{tab:cme_context}
\renewcommand{\arraystretch}{1}
\setlength{\tabcolsep}{6pt}
\small
\begin{tabular}{p{0.12\textwidth} p{0.12\textwidth} p{0.25\textwidth} p{0.20\textwidth}  p{0.22\textwidth}}
\hline
Event date &
Eruption time (UT) &
Source / Location &
limb / halo &
In situ (1 AU) \\
\hline

23 Feb 1997 &
C2 02:30:05 &
CPA 208$^\circ$, width 209$^\circ$ (CDAW) &
Partial halo &
No L1 signature. \\[10pt]

30 Apr 1997 &
C2 04:50:38 &
CPA 84$^\circ$, width 71$^\circ$ (CDAW) &
limb / non-halo &
No L1 signature. \\[10pt]

02 Jun 1998 &
C2 21:06:24 &
CPA 50$^\circ$, width 61$^\circ$ (CDAW) &
limb / non-halo &
No clear 1 AU match. \\[10pt]

11 Jun 1998 &
C2 10:28:38 &
CPA 123$^\circ$, width 177$^\circ$ (CDAW) &
Partial halo &
L1 CME: sheath 1998-06-13 19:18, ejecta 1998-06-13 22:18 -- 1998-06-14 23:26. \\[10pt]

17 May 2008 &
C2 10:36:04 &
$\sim$77$^\circ$ east of ST-A &
Partial halo &
No L1 signature. ST-B in situ.\\[10pt]

03 Apr 2010 &
C2 10:33:58 &
halo (360$^\circ$, CDAW) &
halo &
L1 CME: sheath 2010-04-05 07:54, ejecta 2010-04-05 12:01 -- 2010-04-06 13:20. \\
\hline
\end{tabular}
\end{table*}

Understanding CME evolution through the heliosphere is inherently probabilistic because both the eruptive initial state and the ambient solar wind are only partially constrained. Remote-sensing observations must be converted into three-dimensional initial conditions using geometric and physical assumptions, and are limited by projection effects, finite resolution, and measurement noise. At the same time, CME initiation and early acceleration depend on dynamical processes that vary substantially from event to event and across the solar cycle, while the subsequent propagation is shaped by interactions with a structured, time-dependent heliosphere (e.g., streams, gradients, and other transients). Because the CME--solar-wind coupling is nonlinear, small uncertainties in the initial conditions or background environment can amplify into appreciable divergence in arrival characteristics.

Monte Carlo ensemble modeling provides a direct route to uncertainty propagation by treating uncertain inputs as probability density functions (PDFs) and transforming them, through repeated model realizations, into PDFs for the outputs (e.g., arrival time and leading-edge speed). This is the same basic idea used in probabilistic drag-based approaches, which extend the drag-based model (DBM; \citealt{2013SoPh..285..295V}) by repeatedly sampling uncertain CME and solar-wind inputs to produce distributions of arrival-time and speed predictions rather than a single deterministic solution (e.g., \citealt{2022SpWea..2002925N}). Monte Carlo/ensemble concepts are also used in the evaluation of operational arrival-time forecasts \citep{2018JSWSC...8A..17W}. A key implication is that the forecast spread is only as meaningful as the assumed input PDFs. Accordingly, such probabilistic implementations emphasize the derivation of parameter distributions from CME data rather than adopting purely heuristic uncertainty ranges.

We couple the Monte Carlo methodology to our EFR model. Each realization draws an initial state from observation- and theory-informed distributions spanning the main EFR input-parameter groups, including eruption onset height, footpoint separation $S_f$, aspect ratio $R/a$, magnetic field strengths ($B_p$, $B_t$), the poloidal-flux injection profile, and background solar-wind density and speed. The sampled CME then evolves under Lorentz self-forces (\ref{eq:lorentz}), gravity (\ref{eq:gravity}), aerodynamic drag (\ref{eq:drag}), and sheath pile-up (\ref{eq:mvirtual}), governed by the EFR force-balance (\ref{eq:balance}) and expansion equation (\ref{eq:expansion}). 

For each parameter, we represent the uncertainty with a truncated normal distribution centered on the nominal value, $\mu$, and bounded to $\pm 20\%$, i.e., $[0.8\mu,\,1.2\mu]$. We adopt $\pm 20\%$ as a nominal fractional uncertainty for two practical reasons. First, ensemble CME-arrival studies commonly propagate input uncertainty using prescribed perturbation ranges of this order (roughly $\sim$10--20\%) when assessing forecast sensitivity \citep{2015SpWea..13..611C,2018SpWea..16..784A}. Second, additional uncertainty arises from imperfect specification of the ambient solar wind and from model idealizations. Adopting a modestly conservative fractional envelope therefore provides a consistent baseline for inter-comparison across parameter groups. The truncation limits suppress extreme draws and help preserve physically plausible realizations.

We compute ensembles of $N=1000$ realizations per scenario, which yields stable estimates of distributional quantities (e.g., medians and quantiles) while keeping runtime tractable for the full EFR evolution. We verified that increasing $N$ beyond 1000 does not materially change the reported summary statistics.

Each realization proceeds in two phases:
\begin{enumerate}
    \item Relaxation to a steady state. We first integrate the system through a fixed 2-day stabilization interval, allowing any initialization-driven transients to decay and the background to settle into a consistent starting configuration. In selected experiments, we varied $\tau_{\mathrm{relax}}$ to assess sensitivity and found no material changes in the resulting outputs. Over the final portion of the relaxation interval, key quantities exhibit only weak residual variability (e.g., the velocity changes by $<1\%$), indicating that a quasi-steady initial state has been reached.
    \item Driven EFR evolution. Starting from the relaxed state, we initiate the poloidal-flux injection and evolve the EFR dynamics forward in time under the specified $d\Phi_p/dt(t)$ profile.
\end{enumerate}

\section{Results}

Each Monte Carlo realization yields a complete Sun-to-1~AU evolution of the modeled CME, spanning bulk kinematics and derived quantities relevant to space-weather impact. Across all six events, we evaluated both the time-dependent kinematic profiles and the 1~AU endpoint diagnostics (e.g., arrival time, leading-edge speed, magnetic-field intensities, and impact duration), using the $N=1000$-member ensemble to quantify uncertainty. 

\begin{figure*}[]
    \centering
    \includegraphics[width=16cm]{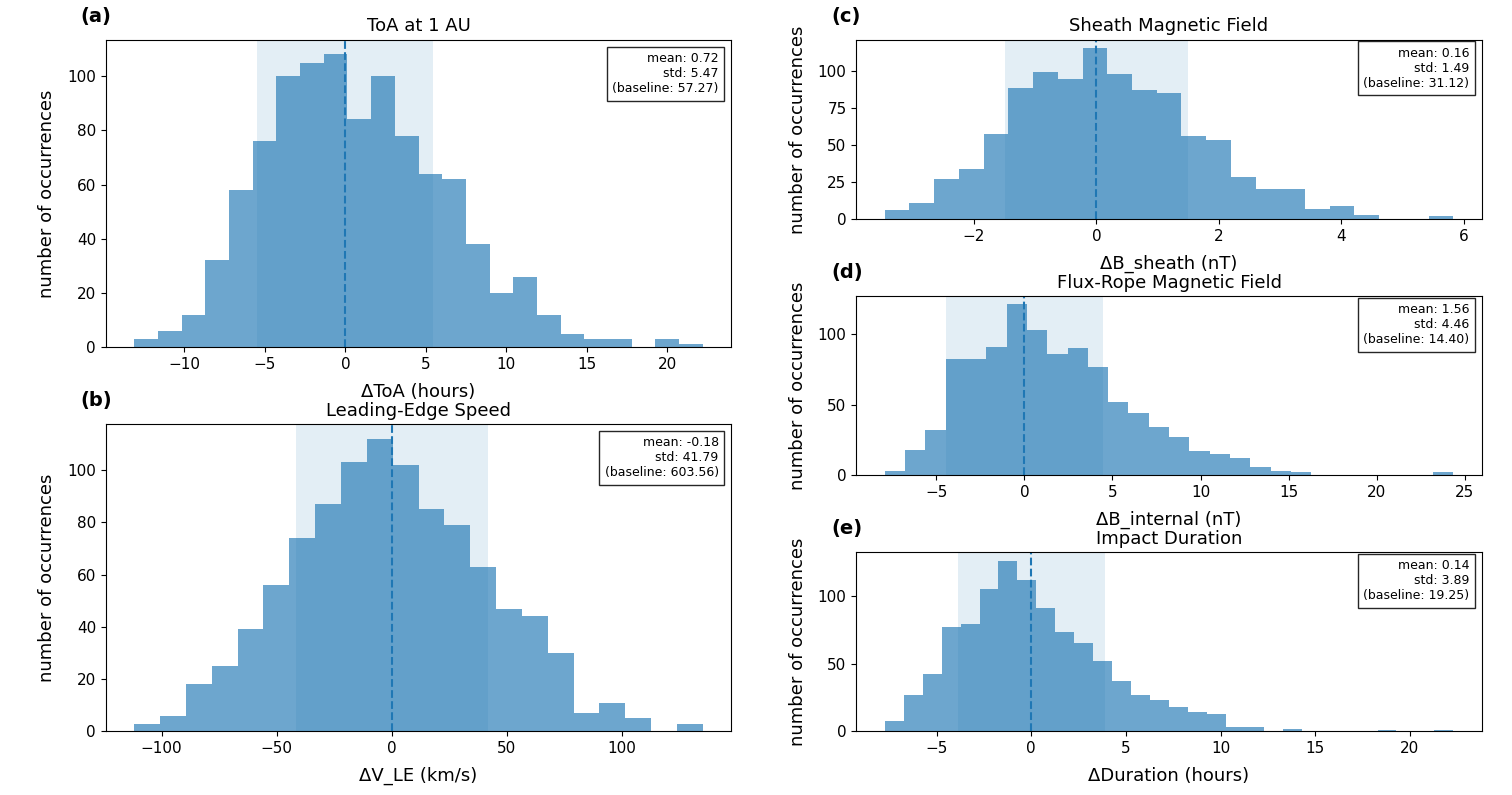}
    \caption{Endpoint spreads at 1~AU for the comprehensive ensemble of the 23 February 1997 event, plotted relative to the standard-input (baseline) solution. Panels show the deviations in (a) time of arrival, (b) leading-edge speed, (c) sheath magnetic-field strength, (d) internal (flux-rope) magnetic-field strength, and (e) impact duration. In each panel, the histogram is annotated with the mean offset, and a faint shaded band indicates the $\pm 1\sigma$ spread. The dashed vertical line marks zero deviation (the standard-input solution).}
    \label{endpoints_23feb1997}
\end{figure*}

The ensemble behavior is summarized using median kinematic profiles with 5th--95th percentile bounds, together with the corresponding distributions of 1~AU diagnostics (Figure~\ref{endpoints_23feb1997}). In the kinematic envelopes, the spread is typically largest in the early propagation phase and then narrows with distance, consistent with progressive momentum coupling to the ambient wind: as the relative CME--solar-wind speed decreases, drag acts to reduce differences in the bulk speed, yielding a tighter band at larger heliocentric distances. This behavior is evident in the representative example shown below.

For the 23 February 1997 event, Figures~\ref{kinematics_23feb1997}--\ref{endpoints_23feb1997} quantify the spread in both the Sun--to--1~AU kinematic evolution and the corresponding 1~AU endpoint diagnostics. Figure~\ref{kinematics_23feb1997} shows the median leading-edge speed profile together with the 5th--95th percentile envelope, while Figure~\ref{endpoints_23feb1997} summarizes the endpoint distributions at 1~AU. 

To assess sensitivity to event-specific initialization, we apply the same Monte Carlo procedure to five additional CME events (30 April 1997, 2 June 1998, and 11 June 1998, 17 May 2008 and 3 April 2010). For each event, we generate an $N=1000$ ensemble and extract the same set of 1~AU diagnostics. The initial parameters for the 11 June 1998 event follow \citet{2010ApJ...717.1105C}, for the 23 February 1997, 30 April 1997, and 2 June 1998 events are adopted from \citet{2003JGRA..108.1410C}, whereas those for 17 May 2008 and 3 April 2010 from \citet{2022JGRA..12728744L}. Throughout this work, these adopted parameter sets define the standard-input solution for each event, against which the Monte Carlo perturbations are applied and overplotted in the kinematic profiles and endpoint distributions. The corresponding kinematic envelopes and endpoint histograms are provided in the Appendices using the conventions of Figures~\ref{kinematics_23feb1997}--\ref{endpoints_23feb1997}. In the following, we summarize the results in compact form.

\begin{figure}[]
    \centering
    \includegraphics[width=\columnwidth]{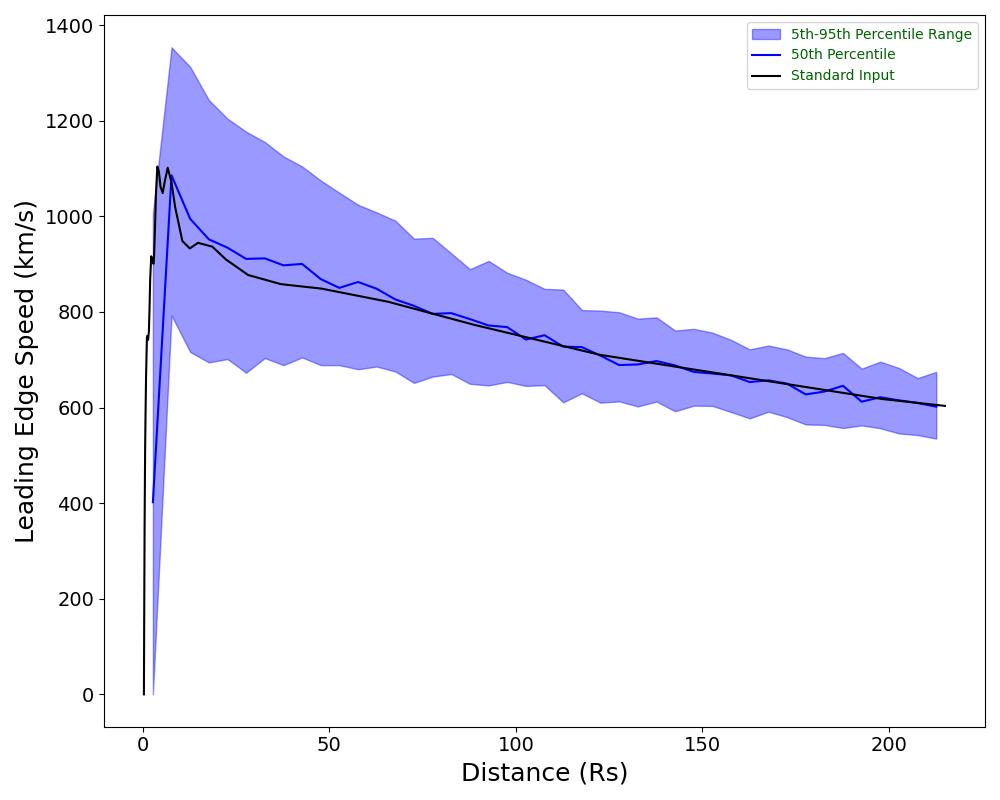}
    \caption{Leading-edge evolution of the CME flux rope as a function of the heliocentric distance for the 23 February 1997 event. The blue shaded region denotes the 5th--95th percentile envelope and the solid blue curve is the median. The standard-input solution is over-plotted in black.}
    \label{kinematics_23feb1997}
\end{figure}

To provide observational context, we also compare two events for which in situ ICME signatures are available in the HELIO4CAST ICME catalog (ICMECAT) \citep{2017SpWea..15..955M}. This comparison is not intended as a formal forecast-skill assessment, since our primary objective is to characterize ensemble dispersion and parameter importance within the EFR framework. Nevertheless, it offers a basic check that the modeled Sun--to--1~AU evolution produces arrival and impact diagnostics of a reasonable order.

For the 11 June 1998 event (Figure~\ref{endpoints_11june1998}), the in situ ToA is 59.83~h, compared to 62.05~h for the standard-input solution and 61.01~h for the ensemble mean, while the in situ impact duration is 25.13~h compared to 23.02~h (standard) and 23.74~h (mean). The in situ peak magnetic field and arrival speed are 12.4~nT and 347.6~km~s$^{-1}$, versus 21.69~nT and 522.56~km~s$^{-1}$ (standard) and 20.17~nT and 549.55~km~s$^{-1}$ (mean), respectively. For the 3 April 2010 event (Figure~\ref{endpoints_03april2010}), the in situ ToA is 49.43~h, compared to 51.64~h (standard) and 53.51~h (mean), and the in situ duration is 28.82~h compared to 12.30~h (standard) and 12.61~h (mean). The in situ peak magnetic field and arrival speed are 20.6~nT and 635.0~km~s$^{-1}$, versus 24.40~nT and 719.13~km~s$^{-1}$ (standard) and 22.20~nT and 746.8~km~s$^{-1}$ (mean).

Overall, these two cases indicate that the modeled arrival timing is captured to within a few hours for the events examined here, while also showing that additional constraints and/or interaction physics may be needed to reproduce event duration and field/speed magnitudes consistently within a single semi-analytic prescription. These comparisons should be interpreted in the regime of relatively direct encounters, where the spacecraft samples the central portion of the magnetic structure (low impact parameter).

To quantify how uncertain inputs affect the modeled variability, we compute correlation-based F scores for each event and each 1~AU diagnostic. Following the feature-ranking procedure of \citet{2018ApJ...855..109L}, we treat each sampled input parameter as a feature and each 1~AU diagnostic as a target. For a given target, we compute the Pearson correlation coefficient $r$ between the target values and each feature across the $N=1000$ realizations, and convert $r$ to an F score,
\begin{equation}
F = \frac{(N-2)\,r^{2}}{1-r^{2}}.
\end{equation}
Because $F$ depends on $r^{2}$, it measures the strength of the linear association and is used here strictly for ranking: larger $F$ indicates a stronger association, while $F\approx 0$ indicates a weak one. Therefore, we report normalized $F$ scores in the tables for comparison between inputs and diagnostics.

Since Pearson correlation is most sensitive to linear relationships, we also compute alternative correlation measures to test the robustness of the rankings. Specifically, we repeat the analysis using the Spearman rank correlation coefficient $\rho$, which captures monotonic (not necessarily linear) relationships, and the distance correlation $\mathrm{dCor}$, which is designed to detect more general forms of dependence: it can be nonzero for nonlinear and nonmonotonic relationships, and it vanishes only when the feature and target are statistically independent. For each metric, we compute the corresponding dependence between each feature and target across the ensemble and derive an analogous score for ranking. Across all events and diagnostics, the rankings obtained using $\rho$ and $\mathrm{dCor}$ are qualitatively consistent with the Pearson-based results, in that the same leading parameters are typically identified. Given this general agreement, we adopt the widely used Pearson-based F scores as the primary importance metric for the results shown below and use Spearman and $\mathrm{dCor}$-based rankings as robustness checks.

\begin{figure*}[]
    \centering
    \includegraphics[width=16cm]{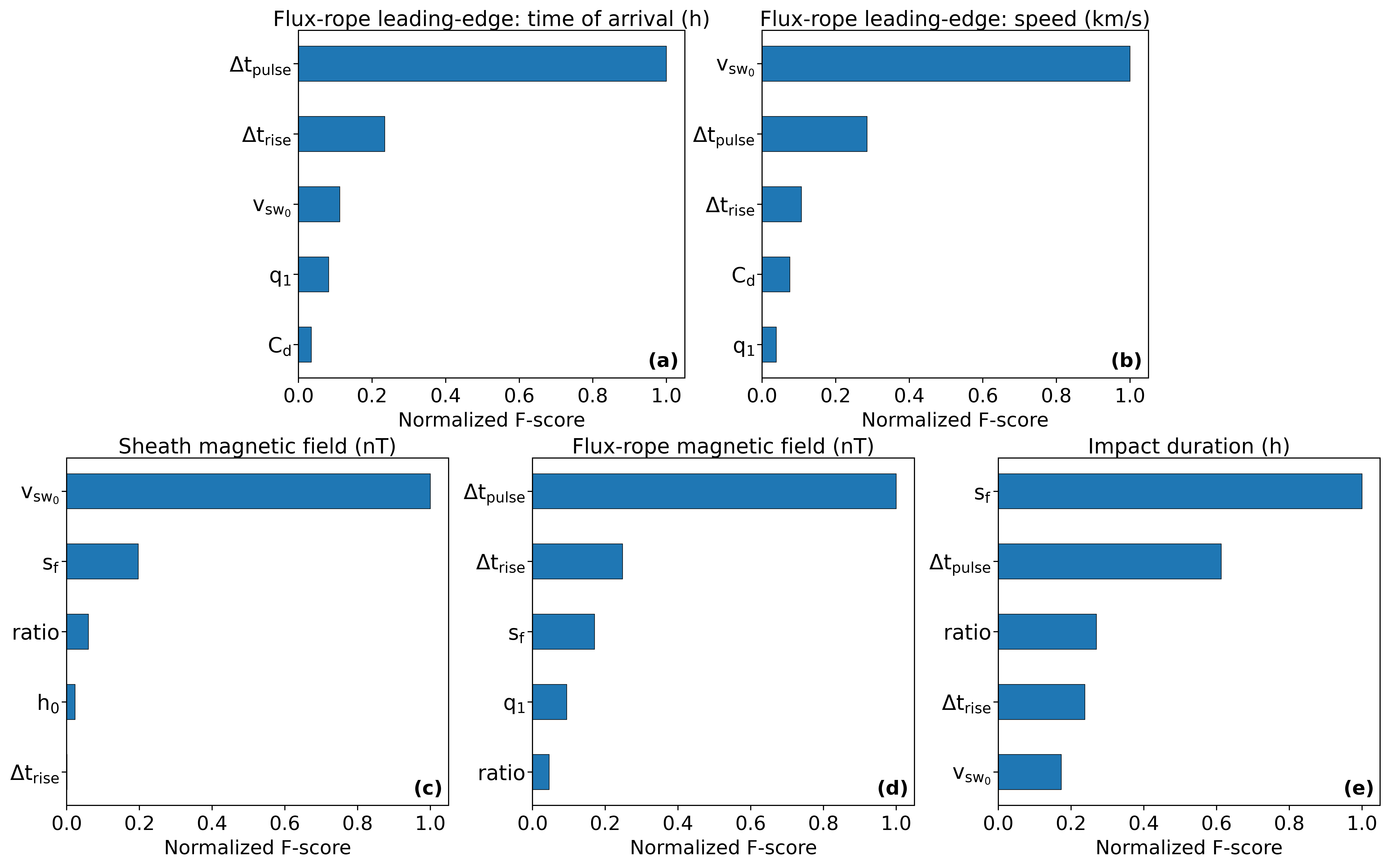}
    \caption{Top five input parameters ranked by the Pearson correlation–based F score for each 1~AU diagnostic in the comprehensive ensemble for the 23 February 1997 event.}
    \label{f_scores_1997-02-23}
\end{figure*}

Tables~\ref{f_scores_1997-02-23} and \ref{f_scores_1997_04_30} through \ref{f_scores_2010_04_03} list the top five input parameters ranked by the Pearson-correlation-based normalized F score for each 1~AU diagnostic across the six ensembles. The resulting rankings consistently highlight two controlling parameter families: (i) the timing and strength of the prescribed poloidal-flux injection profile (equation \ref{injection}) and (ii) the external momentum-coupling terms that govern solar wind interaction ($\mathrm{v}_{\mathrm{sw}_0}$ and the drag coefficient). Geometric scalings (sf, ratio) are comparatively more influential for diagnostics tied to the CME cross-sectional size and passage time at 1~AU (e.g., impact duration).

In the injection profile, $q_{1}$ sets the overall magnitude of the profile by scaling the peak injection amplitude (i.e., the total strength of the imposed poloidal-flux addition). The pulse duration sets the effective width of the injection once it is established, controlling how long the injection remains elevated before declining. The rise duration primarily governs how long the injection builds, whereas the pulse duration governs how long the injection is sustained (the burst/plateau width). Together with $q_{1}$, these parameters encode the onset timing, persistence, and amplitude of the injection profile, and their prominence in the rankings indicates that the modeled 1~AU outcomes are particularly sensitive to the temporal evolution of the driving.

Several robust patterns emerge across the six events. The arrival-time rankings (panel~(a) in Tables~\ref{f_scores_1997-02-23}, and \ref{f_scores_1997_04_30} through \ref{f_scores_2010_04_03}) and the arrival-speed rankings (panel~(b) in the same tables) reflect the same interplay between eruption driving and ambient coupling, but with different emphases. For time of arrival, the dominant contributors are the injected-flux strength $q_1$ and the pulse and rise durations of the injected flux, with the drag coefficient $C_d$ and $\mathrm{v}_{\mathrm{sw}_0}$ acting as recurring secondary terms. Across the additional events shown in the appendices, the same set of top-ranked parameters generally reappears, while the lower-ranked terms reorder among the geometric factors ($ratio$ and $sf$, i.e., the footpoint-separation scaling) and the relaxation timescale $\tau_2$. This pattern indicates that ToA uncertainty in the comprehensive ensembles is organized by both the strength and timing of the early driving and by the subsequent momentum coupling to the background flow. In contrast, the leading-edge speed is consistently governed by the upstream solar-wind speed $\mathrm{v}_{\mathrm{sw}_0}$. For the event discussed here, the next most influential terms are the injection timing ($\Delta t_{\mathrm{rise}}$ and $\Delta t_{\mathrm{rise}}$), followed by the drag coefficient $C_d$ and the injected-flux strength $q_1$. Across the additional events shown in the appendices, the same hierarchy persists, with $\mathrm{v}_{\mathrm{sw}_0}$ remaining dominant while the secondary ordering reshuffles among $C_d$, $q_1$, and the injection-timing terms, with geometric factors ($S_f$ and $ratio$) and $\tau_2$ occasionally entering the top-ranked set.

Sheath-field spread (panel~(c) in the same tables) is dominated by upstream conditions and global scaling, with $\mathrm{v}_{\mathrm{sw}_0}$, $sf$, and $ratio$ comprising the leading terms in most events and $h_0$ and rise duration appearing as lower-ranked contributors. One event departs from this pattern by placing the geometric terms ($S_f$ and $ratio$) above $\mathrm{v}_{\mathrm{sw}_0}$. Internal (flux-rope) magnetic field variability (panel~(d)) is controlled mainly by the injection history together with expansion scaling. Pulse duration (and often rise duration) ranks highly in the main event, while $q_1$ and pulse duration dominate in the additional events, with $sf$ and $ratio$ consistently present and $\tau_2$ (and, in one case, $B_t$) appearing among the remaining top-ranked terms. Duration-related spread (panel~(e)) is organized primarily by geometric scaling, with $sf$ ranking first in all events and $ratio$ consistently appearing next. Secondary contributions reorder among $\mathrm{v}_{\mathrm{sw}_0}$ and driving-related terms (pulse and, in some cases, rise duration), with the initial height $h_0$ and $q_1$ occasionally entering the top set. Together, these rankings provide a concise summary of how inputs affect diagnostics in the ensembles.

\section{Discussion}

The above results provide a baseline quantification of the dispersion in a uniform envelope of input-uncertainty $\pm 20\%$. Here we examine how that choice, together with related model idealizations, influences the ensemble width, and we discuss how improved constraints on the solar wind, eruption driving, and CME geometry could systematically reduce uncertainty in arrival and impact diagnostics.

In our baseline $\pm 20\%$ ensembles, the time-of-arrival (ToA) spread varies across events, spanning $\sim$2.4--7.7~h from the fastest (\ref{11_june_1998}) to the slowest case (\ref{30_april_1997}). The larger spread in the slowest event is consistent with its longer propagation through the structured interplanetary medium and the cumulative influence of CME--wind coupling over extended transit times. However, both spreads remain below the arrival-time uncertainty of order $\sim$10~h summarized by \citet{2019RSPTA.37780096V}. To reproduce an $\sim$10~h ToA uncertainty within our framework, we find that the effective input-error envelope must be increased in an event-dependent manner: for slow events, ramping the perturbation bounds from $\pm 20\%$ to $\pm 25\%$ is sufficient to raise the ToA spread to $\sim$10~h, whereas for fast events the same target spread requires substantially larger perturbations, up to $\pm 40\%$. This contrast highlights that ensemble width is not a universal property of a nominal perturbation level, but depends on both the event regime and how the assumed input distributions reflect real errors and variability.

Across the $\pm 20\%$ perturbation range, variations in the initial magnetic-field strength introduce only modest changes to the CME's initial flux and magnetic pressure. By contrast, the prescribed poloidal-flux injection can add (or withhold) a substantially larger amount of flux during the eruption and directly modifies the Lorentz driving and subsequent expansion. Because this injection-driven evolution reorganizes the acceleration profile and the downstream interaction with the solar wind, it propagates into multiple 1~AU diagnostics, so injection-related parameters tend to outrank the initial-field terms in the F score analysis.

\subsection{Physical interpretation and model limitations}

At the same time, applying the same uncertainty range is not necessarily appropriate. The $\pm 20\%$ envelope provides a consistent baseline, but the most realistic spread is parameter- and event-dependent and should be reduced only when justified by strong observational constraints, for example, geometry and kinematics from multi-viewpoint reconstructions or well-observed early trajectories. On the other hand, background-wind quantities may require broader, and potentially non-Gaussian, uncertainty models to capture structured flows and interaction regions. This distinction is especially relevant for sheath-related diagnostics. The relatively small sheath-field spread in our ensembles likely reflects limited background variability and simplified compression/pile-up representation, rather than implying that sheath conditions are intrinsically well constrained.

These results also highlight the importance of nonlinear coupling in reduced CME dynamics. In the EFR framework, forces, expansion, and momentum exchange remain coupled throughout transit. As a result uncertainty in one input can show up across multiple diagnostics through changes in intermediate variables such as effective inertia, cross-sectional size, and relative CME--wind speed. This is an inherent limitation of semi-analytic modeling: key couplings are represented with simplified, parameterized relationships rather than fully resolved plasma dynamics, and structured-wind interactions may only be captured effectively through variations in bulk background parameters.

We also note a geometric limitation of the present implementation. The EFR model, as used here, does not include CME deflection or rotation during propagation. As a result, our 1~AU magnetic-field and duration diagnostics are most directly interpretable for near-central flux-rope encounters rather than highly off-axis crossings. Differences between events could therefore contribute additional variability in observed in situ profiles that is not captured by the present ensembles.

This limitation is especially important for the southward component $B_z$ at the observer because of its role in geoeffectiveness. In the present framework, however, the spread in $B_z$ at 1~AU cannot be robustly inferred without assumptions beyond the model scope, since it depends not only on the total field strength, but also on the event-dependent flux-rope orientation and spacecraft crossing geometry, and may be further altered by CME rotation during interplanetary propagation. We therefore report the ensemble spread in magnetic-field magnitude, which cannot be directly translated into a meaningful spread in $B_z$, and treat component-level predictions as a natural extension once the orientation and its evolution are explicitly constrained or modeled.

\subsection{Relation to existing ensemble and time-of-arrival forecasting studies}

Our approach also complements the broader CME ensemble literature developed for forecast generation and verification, where the objective is to translate uncertain remote-sensing inputs into a probabilistic arrival-time window and evaluate reliability against in situ outcomes. In this framework, ensemble spread is commonly generated by perturbing CME launch parameters (e.g., speed, width, and direction) and, in some systems, by varying the background solar wind. These studies show that performance depends critically on how input uncertainty is calibrated: under-dispersive ensembles can appear overconfident even when the ensemble-median trajectory is reasonable, whereas overly broad perturbations can reduce utility by inflating false-alarm probability. \citet{2024A&A...689A.187R} demonstrated, using multi-event EUHFORIA cone and spheromak validations, that predicted in situ signatures can depend strongly on both the CME prescription and the ambient solar-wind background. Rather than performing an end-to-end operational skill assessment, we use ensembles as a controlled uncertainty-propagation experiment within an EFR flux-rope dynamical model to diagnose how physically meaningful input variability maps into dispersion across multiple space-weather-relevant outputs, including kinematic and magnetic-structure diagnostics at 1~AU.

It is also useful to place our ensemble ToA spread in the context of reported errors from recent ToA-prediction studies. For example, the CAT-PUMA machine-learning approach reports a mean absolute arrival-time error of $5.9 \pm 4.3$~hr for the test set corresponding to the shuffled split that yielded the highest $R^2$ among their repeated training instances \citep{2018ApJ...855..109L}. This value should therefore be interpreted as an optimized benchmark from their split-selection procedure rather than a fully cross-validated error estimate. A more recent reassessment by \citet{2024ApJ...963..121C} reports a best-split mean absolute error of 7.6 $\pm$ 5.2 hr, while the more conservative cross-validation results yield mean absolute errors above 10 hr. These values are broadly comparable in magnitude to our 2.4--7.7 hr ensemble ToA spread, although the quantities are not directly equivalent: the cited studies report skill against observations, whereas our spread reflects uncertainty propagation within a fixed dynamical model under prescribed input perturbations.

Complementary ensemble strategies have also been developed in reduced-physics frameworks, where computational efficiency enables large sampling and systematic sensitivity studies. \citet{2018ApJ...854..180D} introduced the Drag-Based Ensemble Model (DBEM) as a Monte Carlo extension of drag-based propagation, producing arrival-time and speed distributions from plausible ranges of CME and solar-wind inputs, and \citet{2018JSWSC...8A..11N} showed that probabilistic performance improves when those input distributions are empirically informed rather than selected ad hoc. These drag-based ensembles are closely aligned with our uncertainty-propagation motivation and have the advantage of being readily applied to larger event sets (e.g., the multi-event validation in \citealt{2018ApJ...854..180D}).  While DBEM primarily targets kinematic arrival diagnostics using a prescribed drag law, the EFR-based ensembles retain explicit internal magnetic driving through the poloidal-flux injection history, include expansion physics tied to flux-rope geometry, and incorporate sheath/pile-up and updated drag coupling. This added physics allows us to propagate uncertainty not only into arrival time and arrival speed but also into magnetic-field and duration-related diagnostics at 1~AU, and to attribute the resulting spread to physically interpretable inputs via the feature-ranking analysis. In this sense, even though our present study spans fewer events, it provides a more detailed, model-internal view of how uncertainties in eruption driving, geometry, and ambient coupling map into both kinematic and impact-relevant outputs.

In a closely related imaging-based forecasting context, the constant-speed assumption commonly used in heliospheric-imager time-of-arrival methods had also been relaxed within the ELEvoHI framework, which uses HI-derived kinematics together with drag-based model fitting and ellipse evolution to represent CME propagation through the ambient solar wind \citep{2018SpWea..16..784A}. Along similar lines, \citet{2022SpWea..2003070P} adopted a two-phase kinematic profile, in which the CME accelerates or decelerates in the inner heliosphere and then propagates at an approximately constant speed to 1~AU, yielding improved arrival-time estimates. Together, these approaches emphasize that a substantial fraction of arrival-time uncertainty can originate from imperfect knowledge of the ambient solar wind and its momentum coupling to the CME, and that different diagnostics can be sensitive to different subsets of the input space. Our EFR-based Monte Carlo ensembles adopt the same fast uncertainty-propagation philosophy but target a different level of physics by retaining explicit internal magnetic driving through the poloidal-flux injection history and including sheath/pile-up and drag coupling, enabling dispersion and parameter-importance diagnostics not only for arrival kinematics but also for magnetic-field- and duration-related quantities at 1~AU.

\subsection{Methodological caveats and future observational constraints}

Recent Parker Solar Probe observations motivate the further development of EFR-style approaches. Using WISPR imaging of a slow CME, \citet{2020ApJS..246...72R} developed a 3D EFR-style flux-rope model and found evidence for multi-episode poloidal-flux injection and a substantial role for solar-wind momentum coupling, consistent with the need to represent the CME--wind interaction accurately in reduced models.

Finally, the feature-ranking approach has limitations that should be kept in mind. Pearson-correlation-based F scores emphasize linear associations and can be influenced by inter-feature correlations when multiple inputs have partially redundant effects. We therefore use the rankings as a lightweight screening tool to highlight the dominant inputs for each diagnostic, to be interpreted alongside the ensemble distributions and the known structure of the governing dynamics. In future applications, we will also test the robustness of the inferred ordering using complementary importance measures such as permutation-based rankings and mutual-information-based scores, particularly when nonlinear effects and feature correlations are expected to be strong.

Looking ahead, there are clear and practical ways to tighten this framework by incorporating additional event-specific constraints. Reconnection flux inferred from flare-ribbon evolution of the associated flares can be used to constrain the injected-flux history and link the modeled driving more directly to flare energetics \citep[e.g.,][]{2007ApJ...659..758Q,2017ApJ...845...49K, 2018ApJ...853...41T,2020ApJ...893..141Z}. Preeruptive flux-rope geometry may also be constrained using nonlinear force-free  magnetic field extrapolations \citep[e.g.,][]{2021LRSP...18....1W}. More broadly, applying the same workflow to routine STEREO/SDO-era observations, and to events that also have complementary coverage from Parker Solar Probe or Solar Orbiter together with 1~AU in situ detections, provides a straightforward path toward narrower, better-justified input ranges and, in turn, sharper arrival and impact diagnostics.

\section{Conclusions}

We developed and applied a Monte Carlo uncertainty-quantification framework built around a semi-analytic erupting flux rope model to evaluate how uncertainty in CME and solar-wind inputs propagates into geoeffective-relevant diagnostics at 1~AU. The framework evolves each realization self-consistently from the low corona to 1~AU. In addition to the classic EFR dynamics, we update the solar wind coupling through an improved drag formulation and a sheath/pile-up treatment. For each event, the output is naturally expressed as distributions of arrival time, leading-edge speed, sheath and flux-rope magnetic fields, and impact duration, together with correlation-based F score rankings that identify which inputs most strongly affect the spread in each diagnostic.

For the six analyzed events, the results show that different 1~AU diagnostics are governed by different parts of the input space. Arrival-time sensitivity is event dependent, but the leading contributors repeatedly come from the same small set of parameters describing the injection history and CME--solar wind coupling. The variability of the leading-edge speed is consistently affected by the upstream solar-wind speed, with drag-related coupling contributing in several cases. Magnetic-field diagnostics separate into two regimes: the sheath field remains comparatively tightly distributed and is strongly associated with upstream conditions (or, in one case, with geometric scaling), whereas the internal flux-rope field exhibits larger relative spread and is dominated by poloidal-flux injection parameters, including the injected-flux magnitude and injection-duration terms. Impact-duration variability is dominated by geometric scaling, particularly the footpoint separation, consistent with its control of CME size and expansion.

With inputs sampled within $\pm 20\%$ of their nominal values, the arrival-time uncertainty spans 2.4--7.7~h. More generally, these results show how a fast, physics-informed flux-rope model can translate input uncertainty into quantitative confidence bounds at 1~AU while remaining physically interpretable. The ensemble distributions quantify the spread expected under the assumed input uncertainty, and the feature rankings provide a compact way to relate that spread to the inputs that most strongly influence each diagnostic. Our use of F score ranking follows well-established practice in feature-importance and feature-selection problems, where such scores are commonly used to identify the most informative predictors, not only in CME arrival-time prediction studies \citep{2018ApJ...855..109L, 2024ApJ...963..121C} but also in related space-weather applications such as solar-flare prediction \citep{2015ApJ...798..135B}.

These conclusions establish a practical baseline for uncertainty-aware CME prediction within an extended EFR framework. In the next section, we place this baseline in context by discussing how the inferred ensemble widths depend on the adopted uncertainty envelope and model idealizations, and we outline the most direct pathways for tightening the bounds through improved background specification and observationally constrained driving.

\begin{acknowledgements}
The authors thank the referee for useful comments/suggestions.
S.S. and S.P. acknowledge support by the ERC Synergy Grant 'Whole Sun' (GAN: 810218). E.P. and A.V are supported by NASA grants 80NSSC22K0970 and 80NSSC24K0847.
\end{acknowledgements}

\bibliographystyle{aa-note} 
\bibliography{msref}

\onecolumn

\appendix
\section{30 April 1997 event: comprehensive ensemble results}
\label{30_april_1997}

This appendix presents the Monte Carlo ensemble results for the 30 April 1997 event. We show (i) the kinematic evolution of the flux-rope leading-edge speed as a function of heliocentric distance and (ii) the endpoint distributions at 1~AU for the time of arrival (a), leading-edge speed (b), sheath magnetic-field intensity (c), flux-rope axial magnetic-field intensity (d), and impact duration (e). The plotting conventions match those used for the representative event in the main text (Figures~\ref{kinematics_23feb1997}--\ref{endpoints_23feb1997}).

\begin{figure*}[h]
    \centering
    \includegraphics[width=0.5\columnwidth]{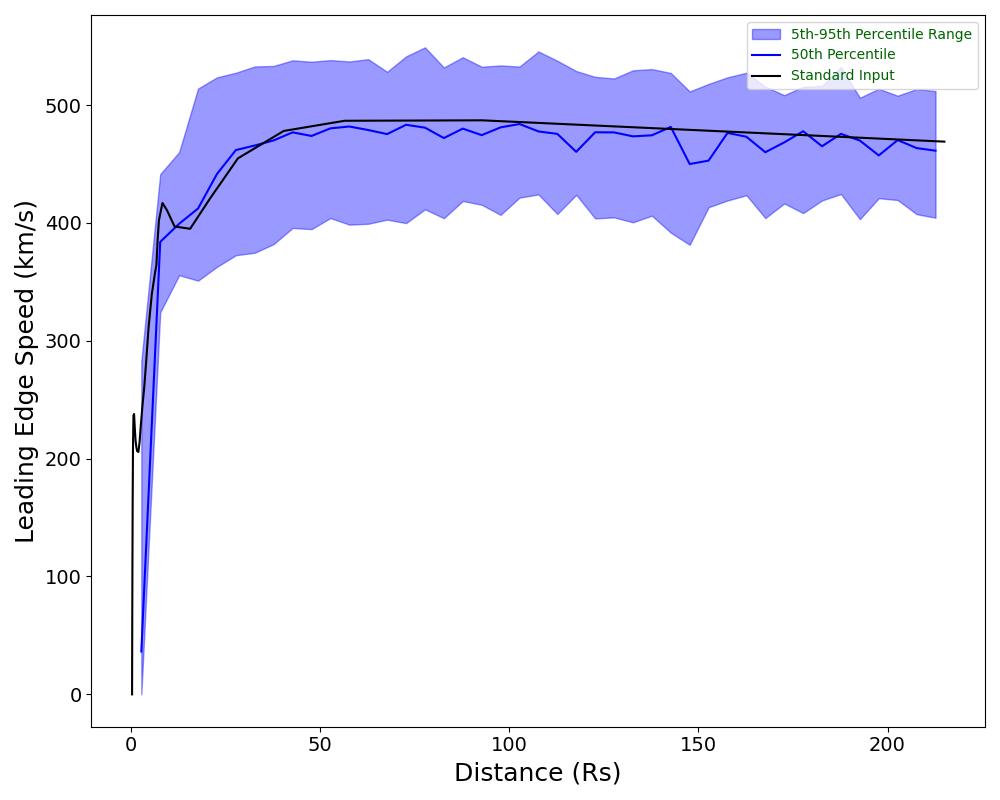}
    \caption{Leading-edge speed of the CME flux rope as a function of heliocentric distance for the 30 April 1997 event. The blue shaded region denotes the 5th--95th percentile envelope, and the solid blue curve shows the median. The standard-input solution is overplotted in black.}
    \label{kinematics_30apr1997}
\end{figure*}

\begin{figure*}[h]
    \centering
    \includegraphics[width=0.85\columnwidth]{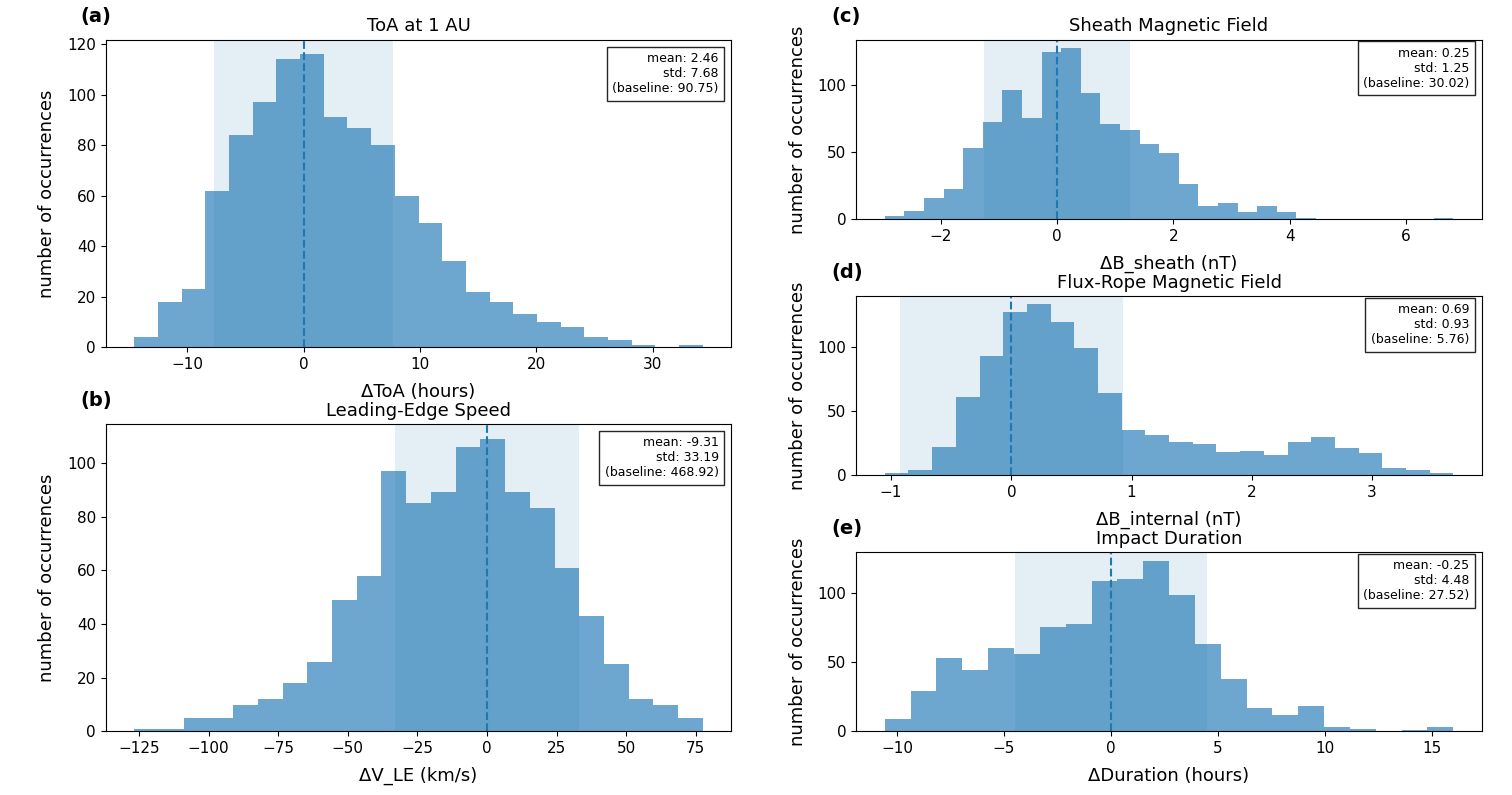}
    \caption{Endpoint spreads at 1~AU for the comprehensive ensemble of the 30 April 1997 event, plotted relative to the standard-input (baseline) solution. Panels show the deviations in (a) time of arrival, (b) leading-edge speed, (c) sheath magnetic-field strength, (d) internal (flux-rope) magnetic-field strength, and (e) impact duration. In each panel, the histogram is annotated with the mean offset, and a faint shaded band indicates the $\pm 1\sigma$ spread. The dashed vertical line marks zero deviation (the standard-input solution).}
    \label{endpoints_30apr1997}
\end{figure*}

\begin{figure*}[h]
    \centering
    \includegraphics[width=0.85\columnwidth]{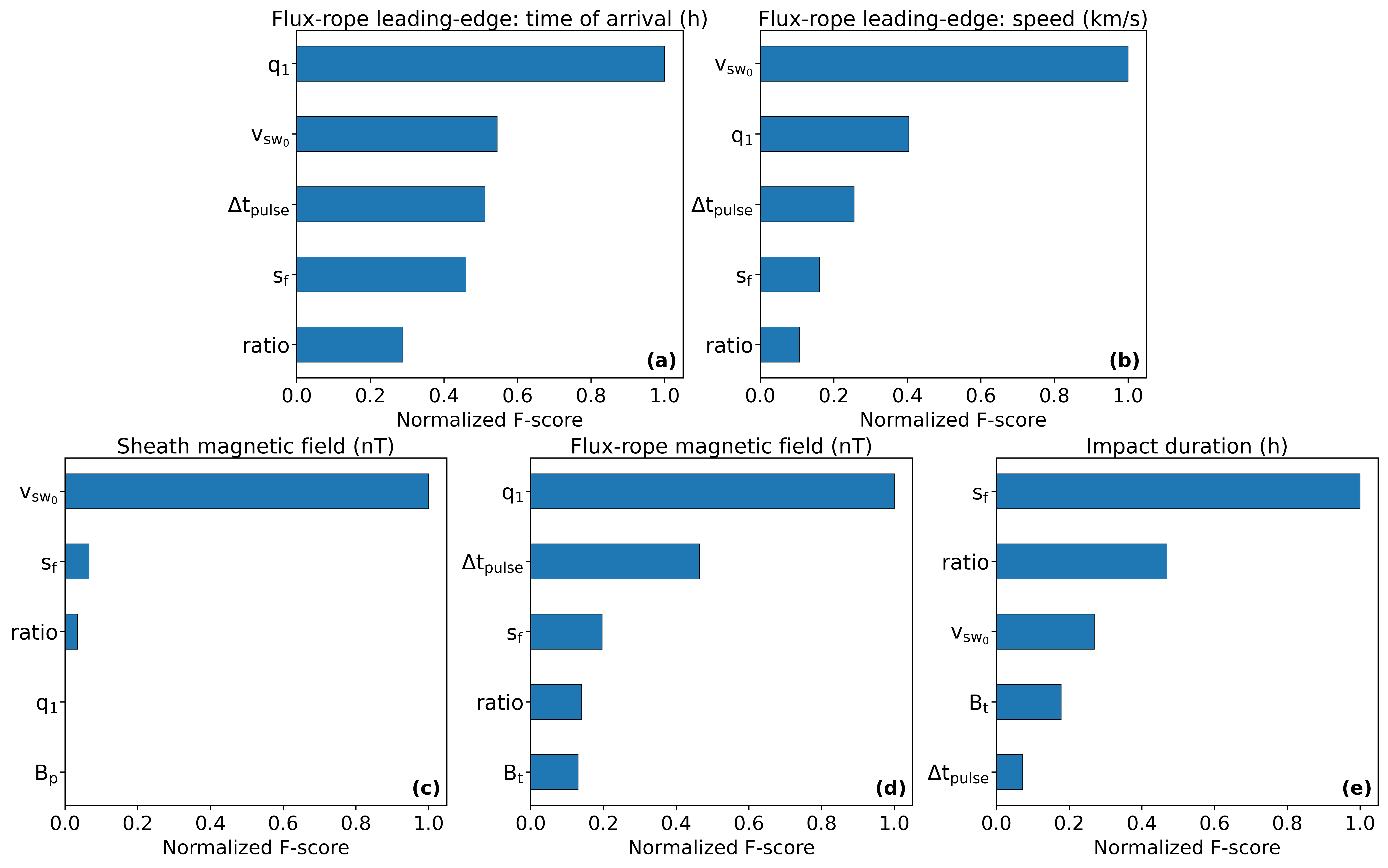}
    \caption{Top five input parameters ranked by Pearson-correlation-based F score for each 1~AU diagnostic in the comprehensive ensemble for the 30 April 1997 event.}
    \label{f_scores_1997_04_30}
\end{figure*}

\section{2 June 1998 event: comprehensive ensemble results}
\label{2_june_1998}
This appendix presents the Monte Carlo ensemble results for the 2 June 1998 event, using the same plotting conventions as in the main text (Figures~\ref{kinematics_23feb1997}--\ref{endpoints_23feb1997}).

\begin{figure*}[h]
    \centering
    \includegraphics[width=0.5\columnwidth]{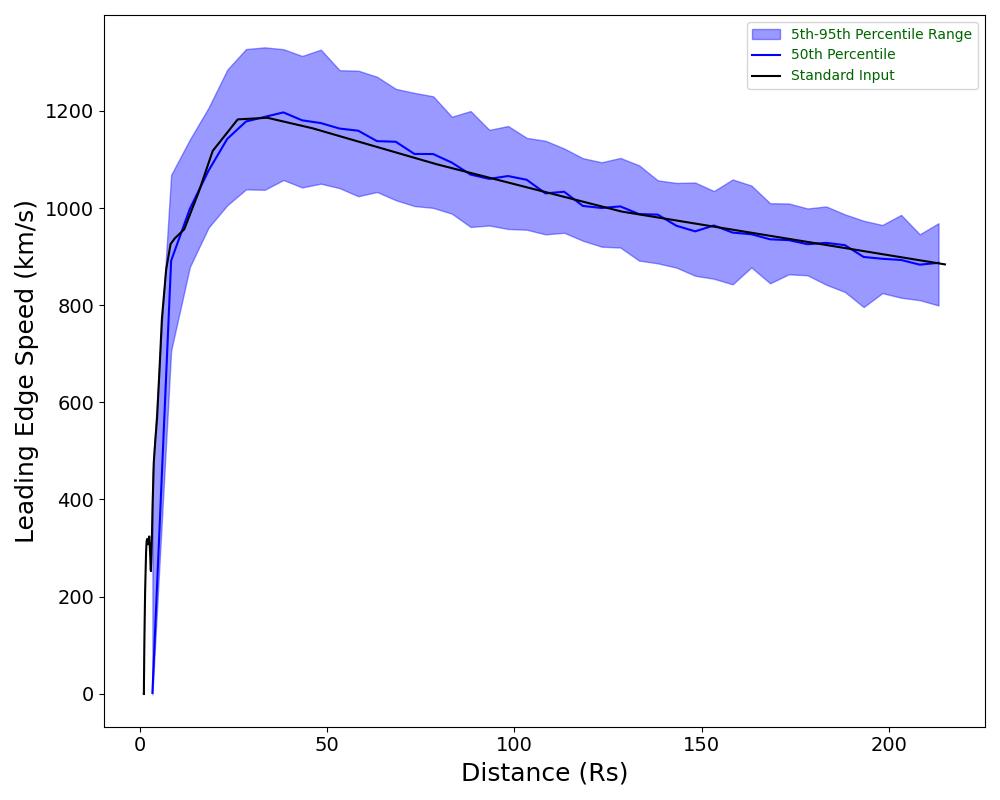}
    \caption{Leading-edge speed of the CME flux rope as a function of heliocentric distance for the 2 June 1998 event. The blue shaded region denotes the 5th--95th percentile envelope, and the solid blue curve shows the median. The standard-input solution is overplotted in black.}
    \label{kinematics_02jun1998}
\end{figure*}

\begin{figure*}[h]
    \centering
    \includegraphics[width=0.85\columnwidth]{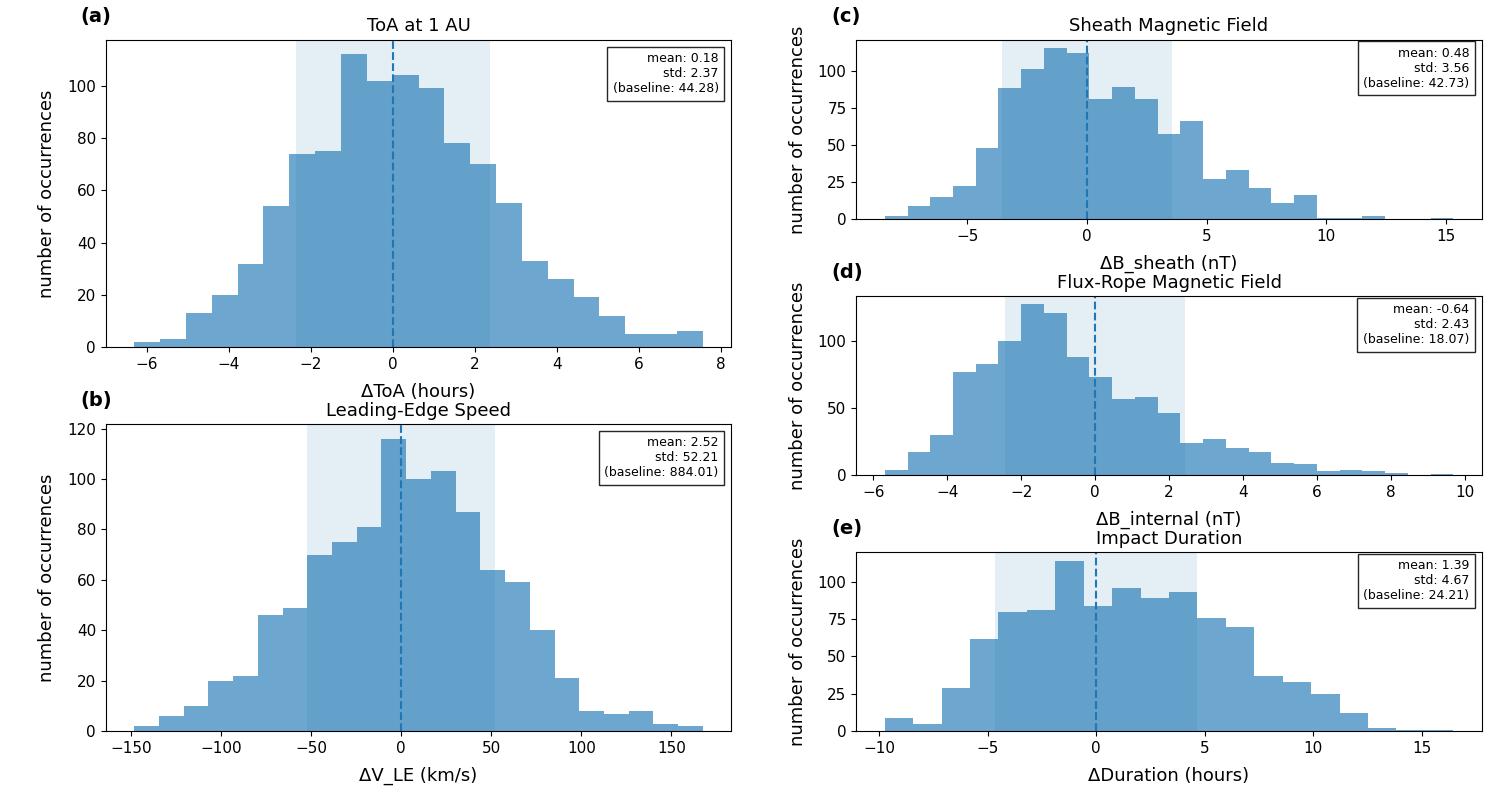}
    \caption{Endpoint spreads at 1~AU for the comprehensive ensemble of the 2 June 1998 event, plotted relative to the standard-input (baseline) solution. Panels show the deviations in (a) time of arrival, (b) leading-edge speed, (c) sheath magnetic-field strength, (d) internal (flux-rope) magnetic-field strength, and (e) impact duration. In each panel, the histogram is annotated with the mean offset, and a faint shaded band indicates the $\pm 1\sigma$ spread. The dashed vertical line marks zero deviation (the standard-input solution).}
    \label{endpoints_02jun1998}
\end{figure*}

\begin{figure*}[h]
    \centering
    \includegraphics[width=0.85\columnwidth]{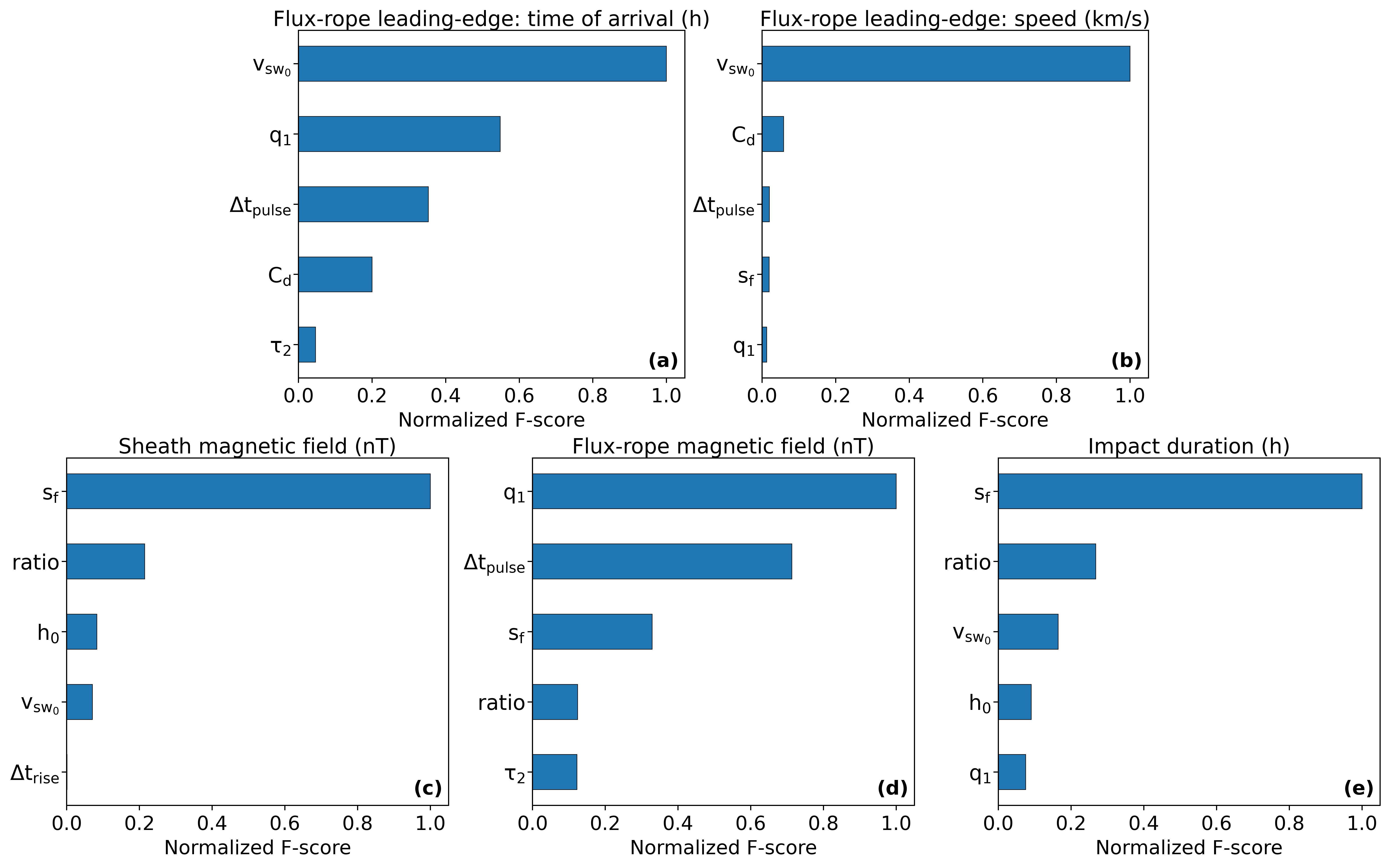}
    \caption{Top five input parameters ranked by Pearson-correlation-based F score for each 1~AU diagnostic in the comprehensive ensemble for the 2 June 1998 event.}
    \label{f_scores_1998_06_02}
\end{figure*}

\clearpage

\section{11 June 1998 event: comprehensive ensemble results}
\label{11_june_1998}

This appendix presents the Monte Carlo ensemble results for the 11 June 1998 event, using the same plotting conventions as in the main text (Figures~\ref{kinematics_23feb1997}--\ref{endpoints_23feb1997}).

\begin{figure*}[h]
    \centering
    \includegraphics[width=0.5\columnwidth]{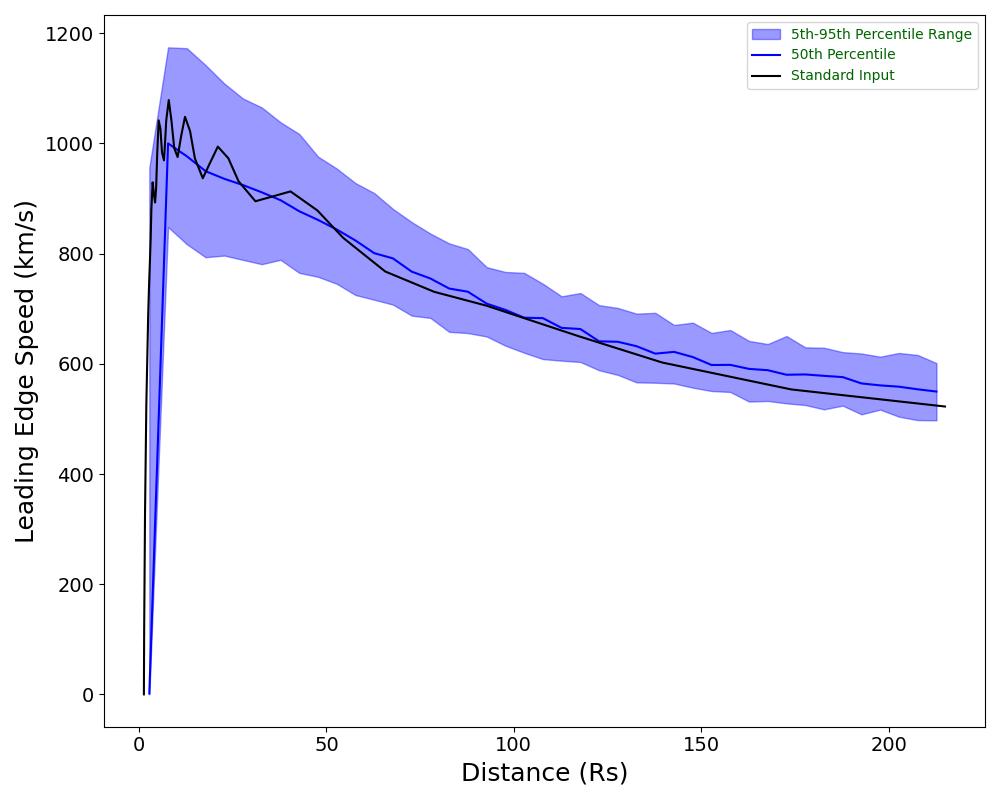}
    \caption{Leading-edge speed of the CME flux rope as a function of heliocentric distance for the 11 June 1998 event. The blue shaded region denotes the 5th--95th percentile envelope, and the solid blue curve shows the median. The standard-input solution is overplotted in black.}
    \label{kinematics_11june1998}
\end{figure*}

\begin{figure*}[h]
    \centering
    \includegraphics[width=0.85\columnwidth]{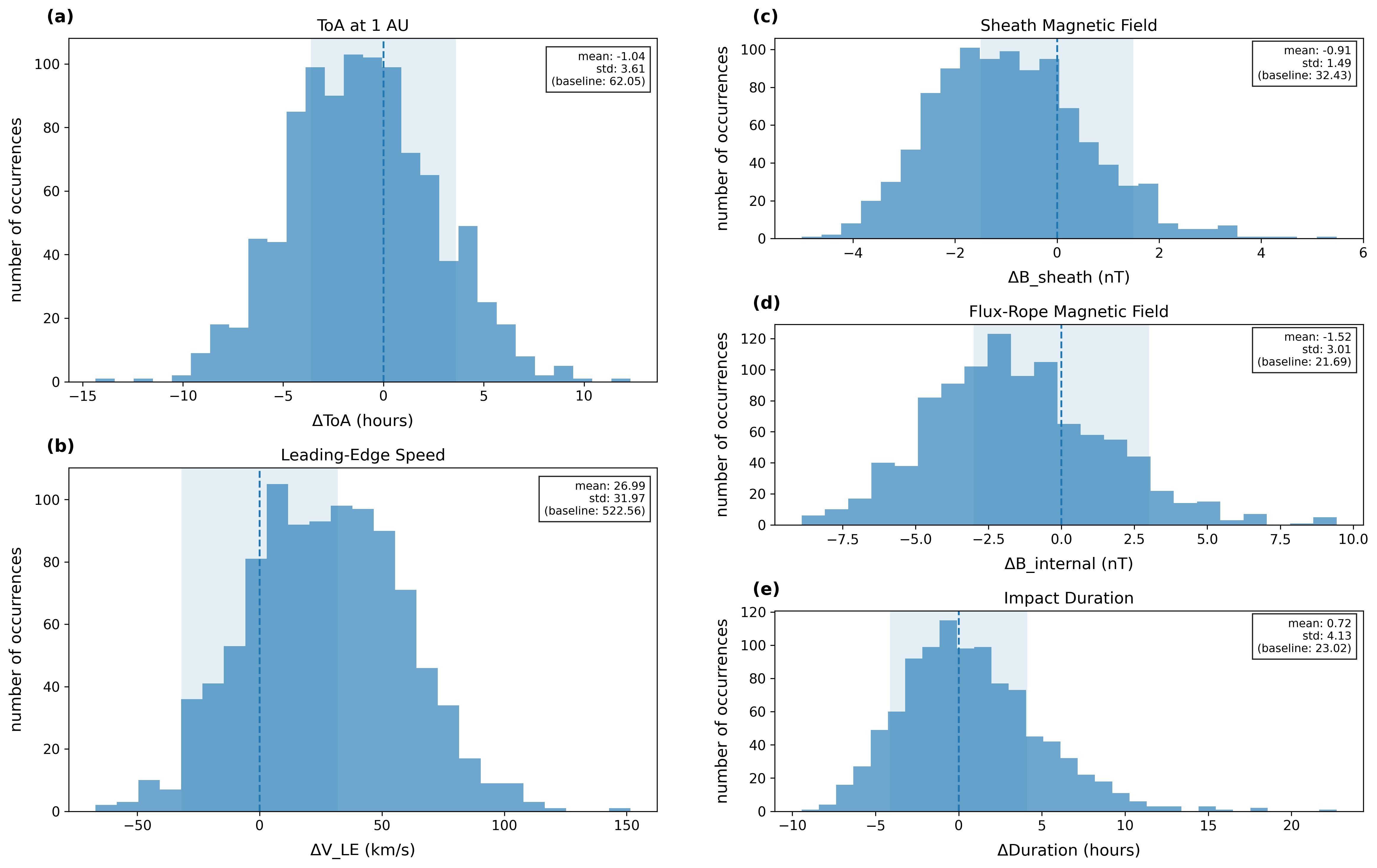}
    \caption{Endpoint spreads at 1~AU for the comprehensive ensemble of the 11 June 1998 event, plotted relative to the standard-input (baseline) solution. Panels show the deviations in (a) time of arrival, (b) leading-edge speed, (c) sheath magnetic-field strength, (d) internal (flux-rope) magnetic-field strength, and (e) impact duration. In each panel, the histogram is annotated with the mean offset, and a faint shaded band indicates the $\pm 1\sigma$ spread. The dashed vertical line marks zero deviation (the standard-input solution).}
    \label{endpoints_11june1998}
\end{figure*}

\begin{figure*}[h]
    \centering
    \includegraphics[width=0.85\columnwidth]{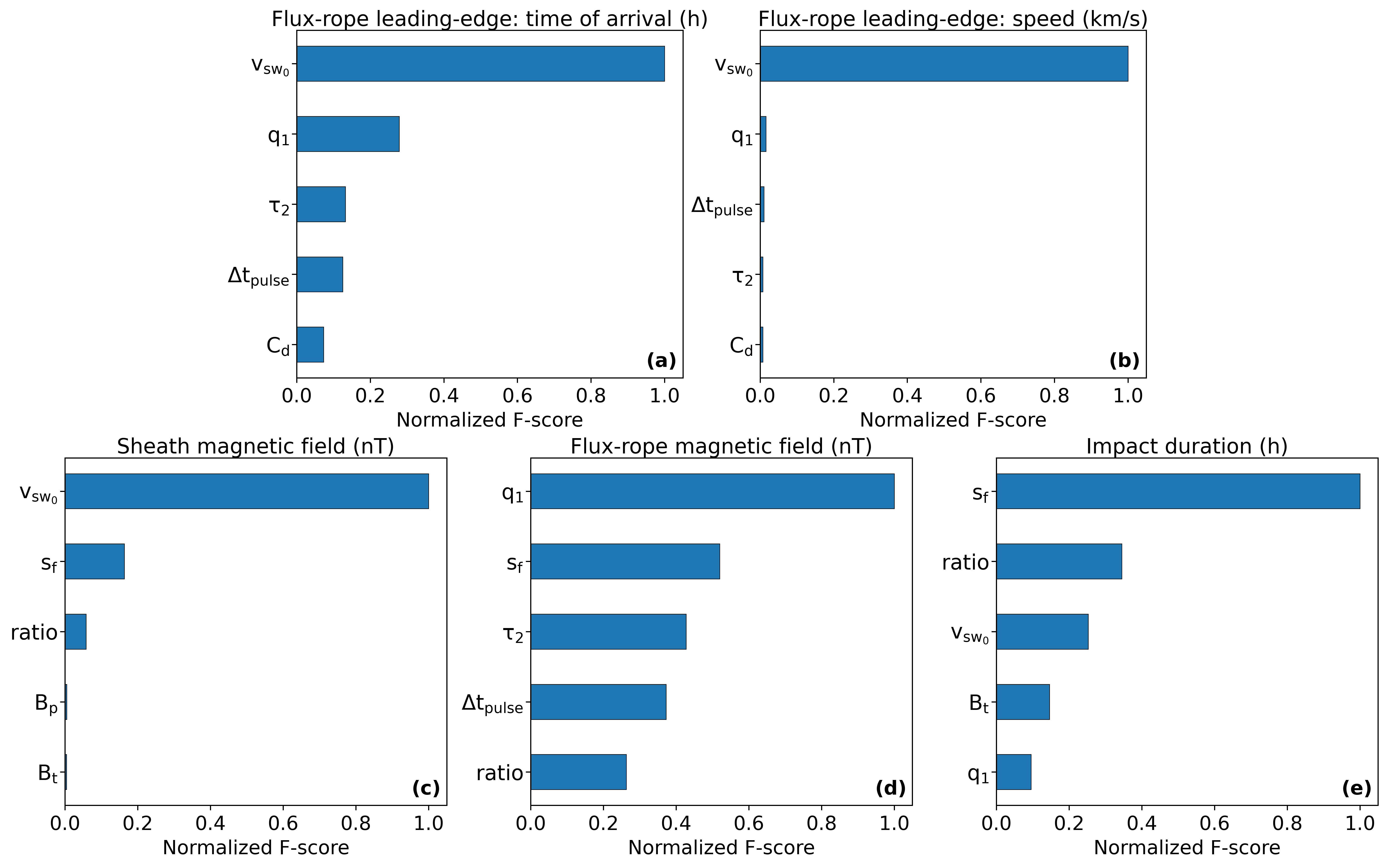}
    \caption{Top five input parameters ranked by Pearson-correlation-based F score for each 1~AU diagnostic in the comprehensive ensemble for the 11 June 1998 event.}
    \label{f_scores_2000_09_12}
\end{figure*}

\section{17 May 2008 event: comprehensive ensemble results}
\label{17_may_2008}

This appendix presents the Monte Carlo ensemble results for the 17 May 2008 event, using the same plotting conventions as in the main text (Figures~\ref{kinematics_23feb1997}--\ref{endpoints_23feb1997}).

\begin{figure*}[h]
    \centering
    \includegraphics[width=0.5\columnwidth]{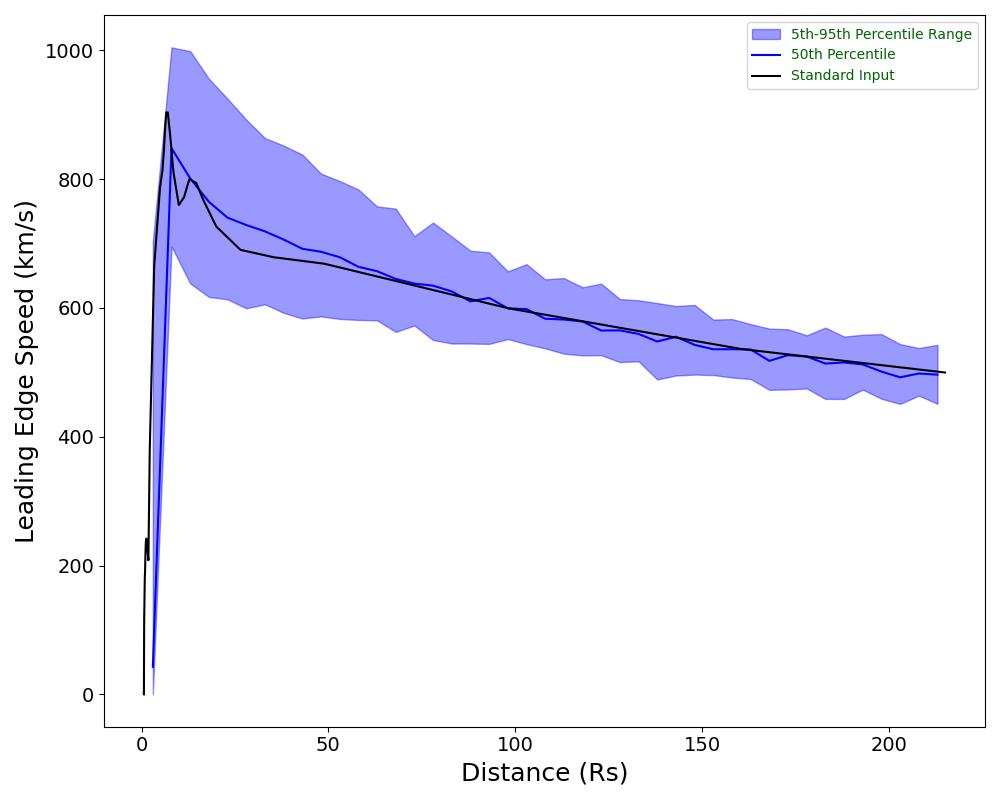}
    \caption{Leading-edge speed of the CME flux rope as a function of heliocentric distance for the 17 May 2008 event. The blue shaded region denotes the 5th--95th percentile envelope, and the solid blue curve shows the median. The standard-input solution is overplotted in black.}
    \label{kinematics_17may2008}
\end{figure*}

\begin{figure*}[h]
    \centering
    \includegraphics[width=0.85\columnwidth]{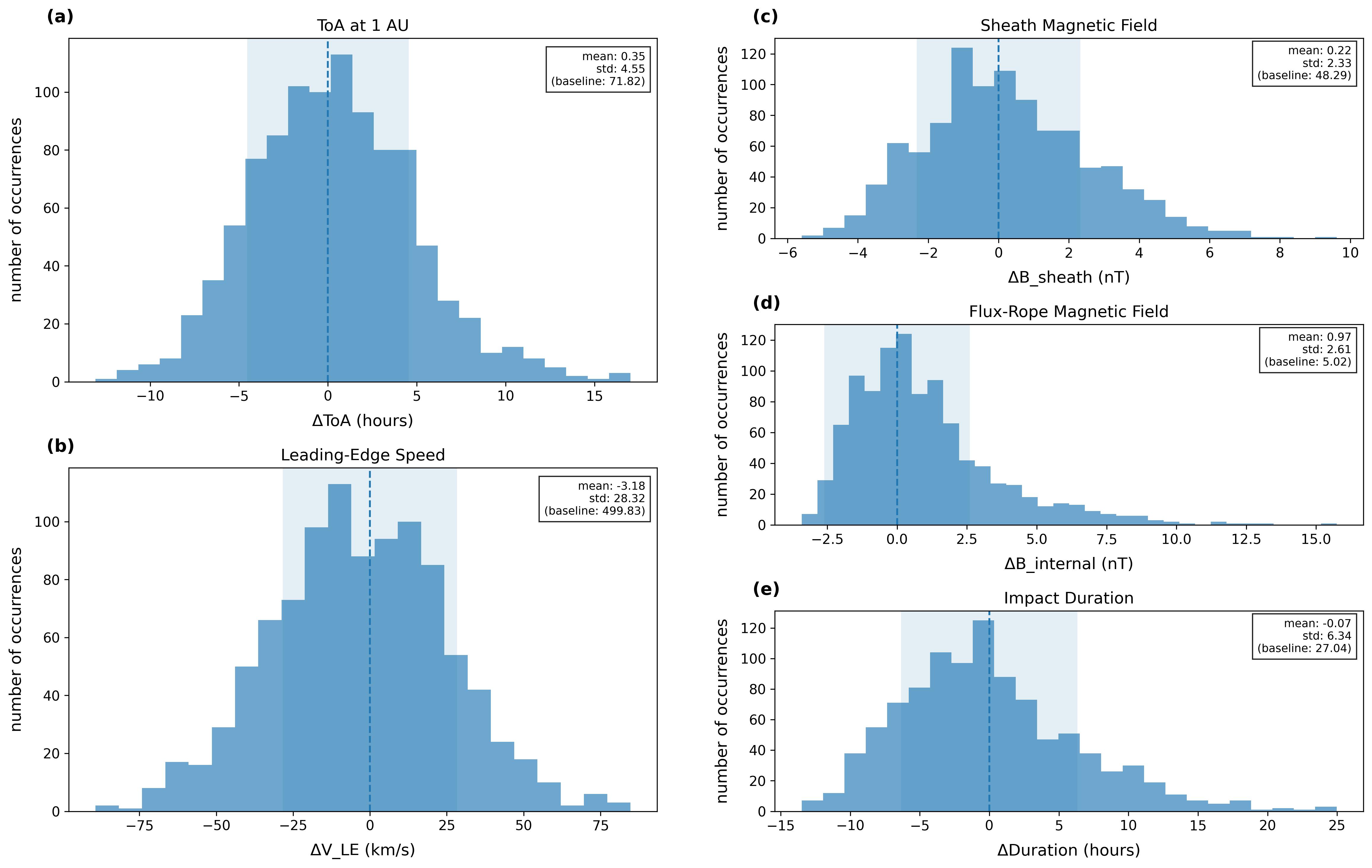}
    \caption{Endpoint spreads at 1~AU for the comprehensive ensemble of the 17 May 2008 event, plotted relative to the standard-input (baseline) solution. Panels show the deviations in (a) time of arrival, (b) leading-edge speed, (c) sheath magnetic-field strength, (d) internal (flux-rope) magnetic-field strength, and (e) impact duration. In each panel, the histogram is annotated with the mean offset, and a faint shaded band indicates the $\pm 1\sigma$ spread. The dashed vertical line marks zero deviation (the standard-input solution).}
    \label{endpoints_17may2008}
\end{figure*}

\begin{figure*}[h]
    \centering
    \includegraphics[width=0.85\columnwidth]{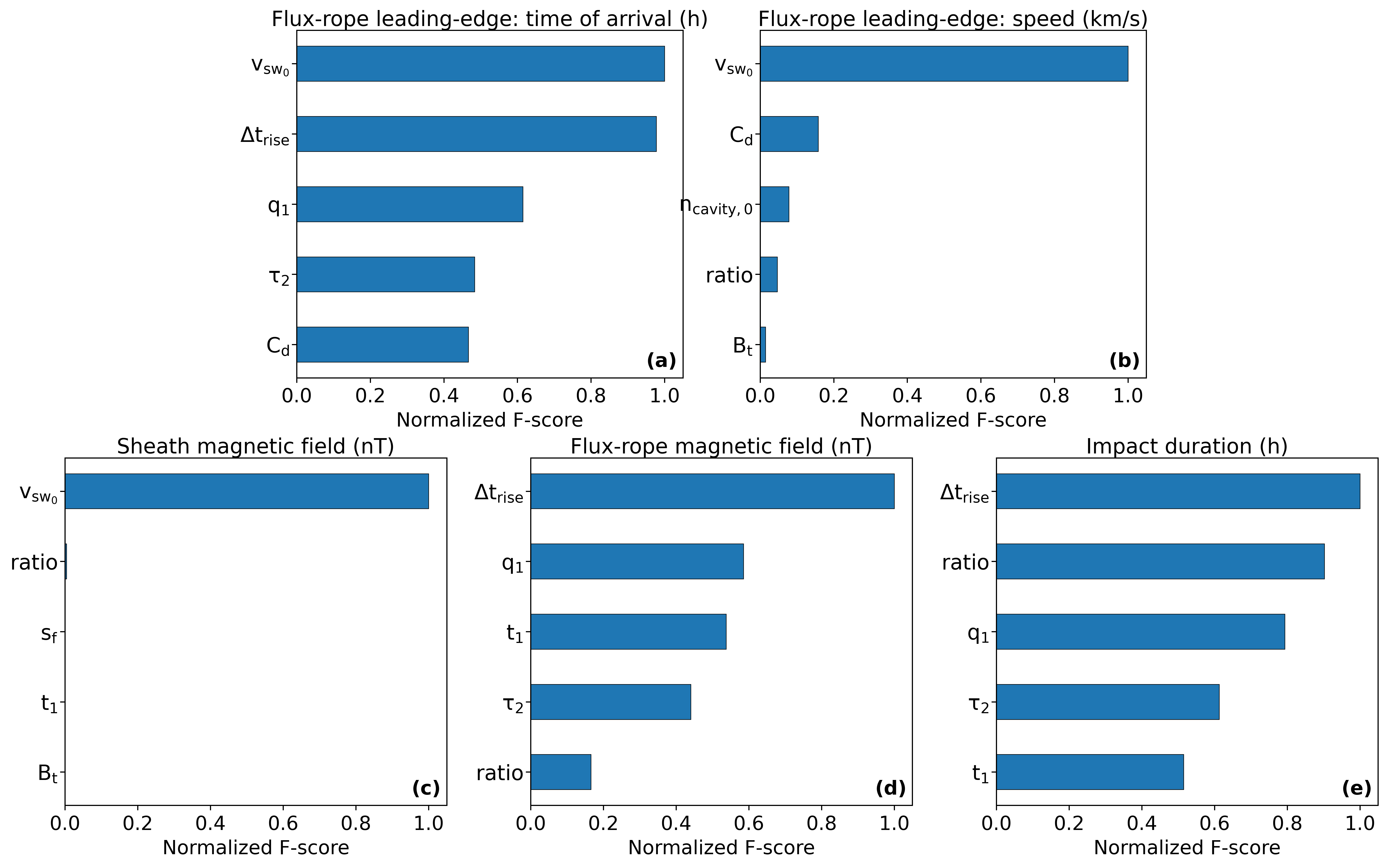}
    \caption{Top five input parameters ranked by Pearson-correlation-based F score for each 1~AU diagnostic in the comprehensive ensemble for the 17 May 2008 event.}
    \label{f_scores_2008_05_17}
\end{figure*}

\section{3 April 2010 event: comprehensive ensemble results}
\label{3_april_2010}

This appendix presents the Monte Carlo ensemble results for the 3 April 2010 event, using the same plotting conventions as in the main text (Figures~\ref{kinematics_23feb1997}--\ref{endpoints_23feb1997}).

\begin{figure*}[h]
    \centering
    \includegraphics[width=0.5\columnwidth]{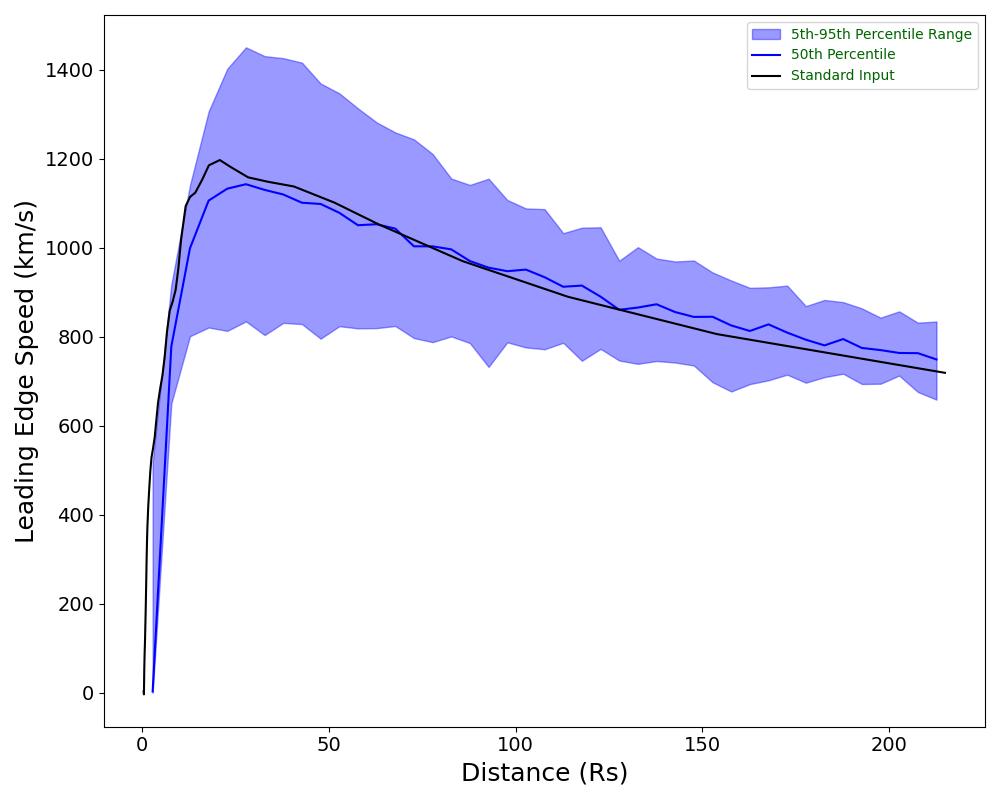}
    \caption{Leading-edge speed of the CME flux rope as a function of heliocentric distance for the 3 April 2010 event. The blue shaded region denotes the 5th--95th percentile envelope, and the solid blue curve shows the median. The standard-input solution is overplotted in black.}
    \label{kinematics_03apr2010}
\end{figure*}

\begin{figure*}[h]
    \centering
    \includegraphics[width=0.85\columnwidth]{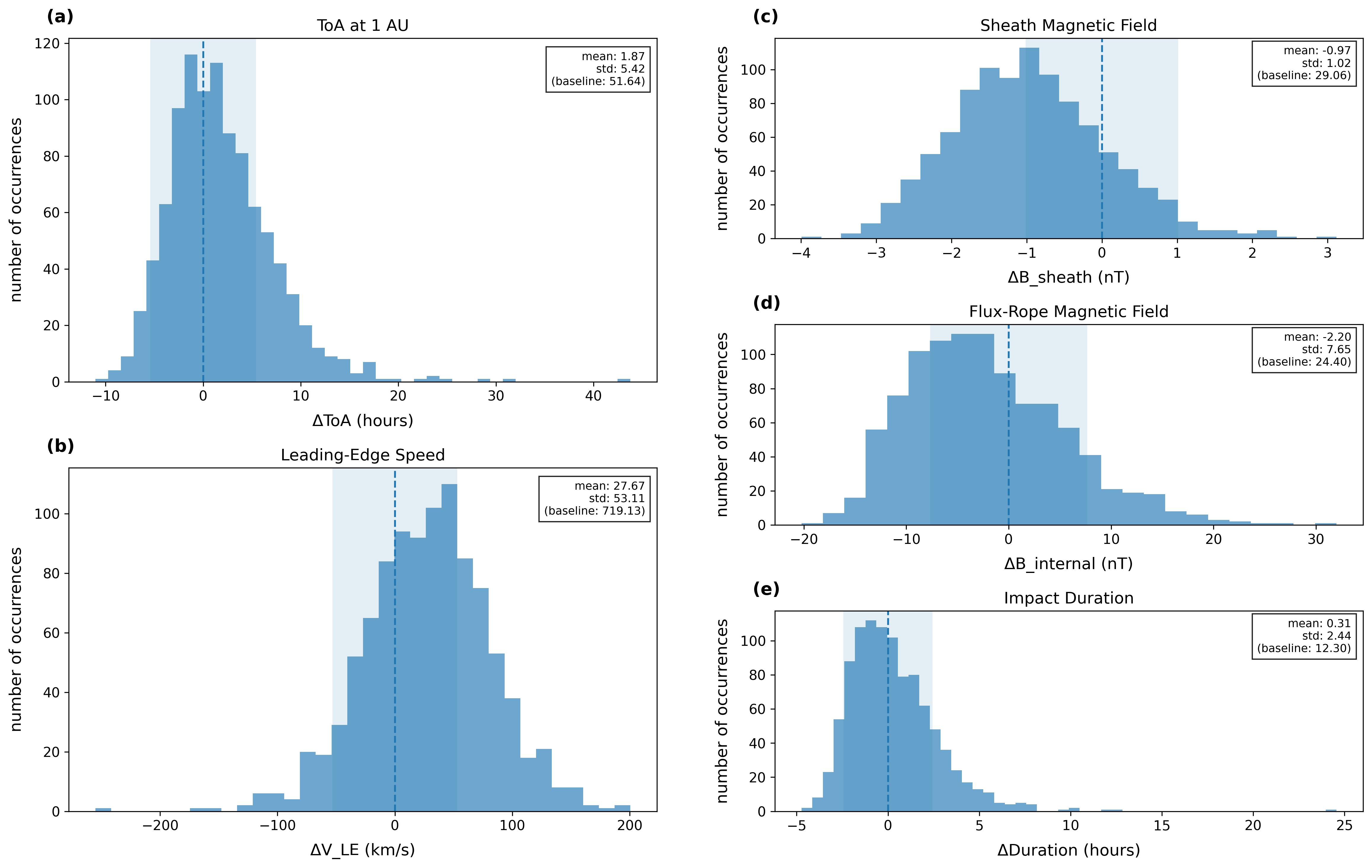}
    \caption{Endpoint spreads at 1~AU for the comprehensive ensemble of the 3 April 2010 event, plotted relative to the standard-input (baseline) solution. Panels show the deviations in (a) time of arrival, (b) leading-edge speed, (c) sheath magnetic-field strength, (d) internal (flux-rope) magnetic-field strength, and (e) impact duration. In each panel, the histogram is annotated with the mean offset, and a faint shaded band indicates the $\pm 1\sigma$ spread. The dashed vertical line marks zero deviation (the standard-input solution).}
    \label{endpoints_03april2010}
\end{figure*}

\begin{figure*}[h]
    \centering
    \includegraphics[width=0.85\columnwidth]{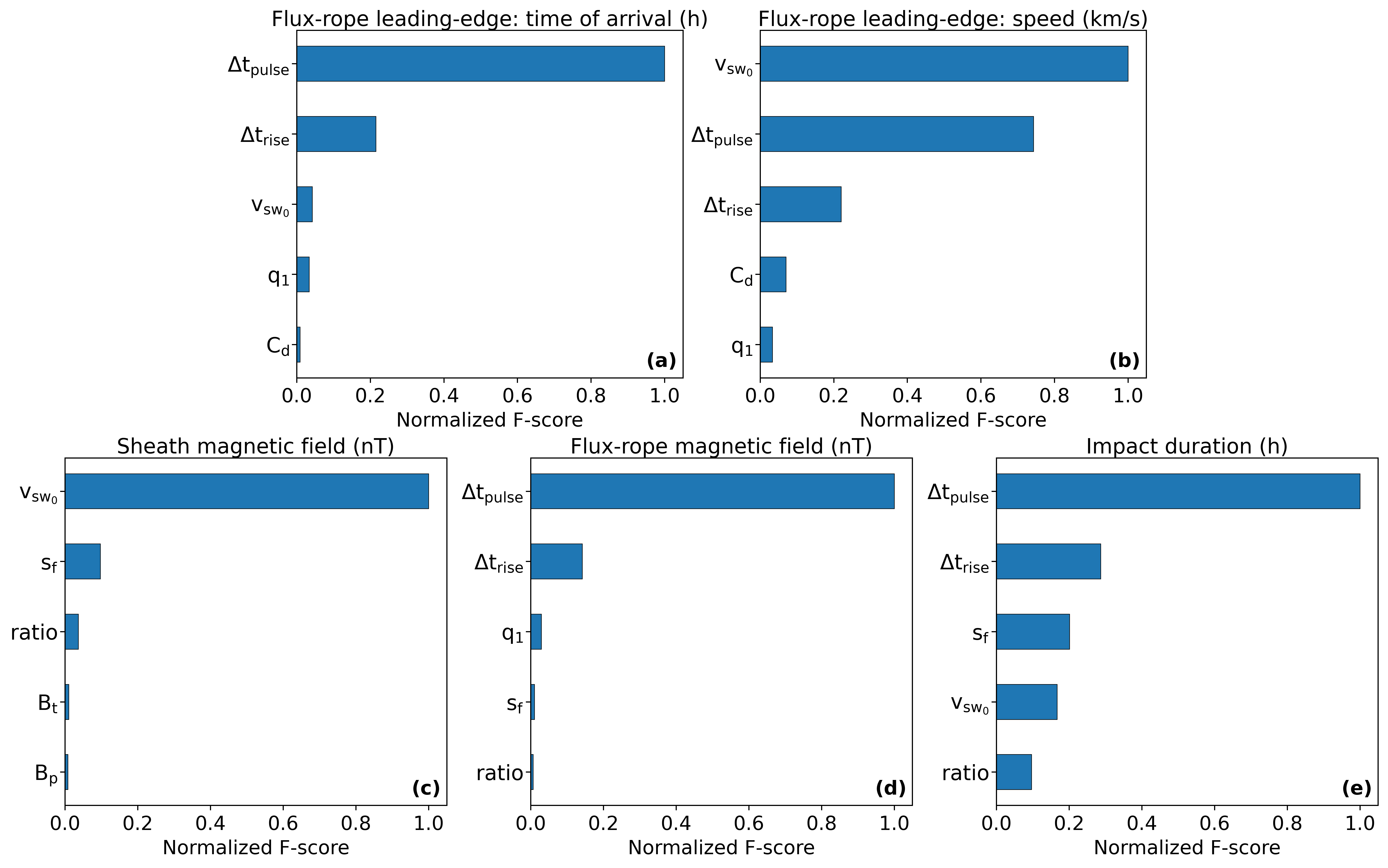}
    \caption{Top five input parameters ranked by Pearson-correlation-based F score for each 1~AU diagnostic in the comprehensive ensemble for the 3 April 2010 event.}
    \label{f_scores_2010_04_03}
\end{figure*}

\end{document}